\documentclass[12pt, draftclsnofoot, onecolumn]{IEEEtran}
\usepackage{graphicx,amsmath,amssymb}
\usepackage{subfigure}
\usepackage{citesort}
\usepackage{fancyhdr}
\usepackage{mdwmath}
\usepackage{mdwtab}
\usepackage{balance}
\usepackage{xcolor}
\usepackage{bm}
\usepackage{amsthm}
\usepackage{algorithm}
\usepackage{algorithmic}
\usepackage{multirow}
\usepackage{flafter}
\usepackage{setspace}
\usepackage{cite}
\usepackage{epstopdf}

\newtheorem{remark}{Remark}

\newtheorem{theorem}{Theorem}

\newtheorem{lemma}{Lemma}

\newtheorem{corollary}{Corollary}

\newtheorem{assumption}{Assumption}

\DeclareMathOperator\erf{erf}
\begin{document}
\title{STAR-IOS Aided NOMA Networks: Channel Model Approximation and Performance Analysis}

\author{Chao~Zhang,~\IEEEmembership{Student Member,~IEEE,}
        Wenqiang~Yi,~\IEEEmembership{Member,~IEEE,}
        Yuanwei~Liu,~\IEEEmembership{Senior Member,~IEEE,}
        Zhiguo~Ding,~\IEEEmembership{Fellow,~IEEE,}
        and Lingyang~Song,~\IEEEmembership{Fellow,~IEEE}
\thanks{C. Zhang, W. Yi, and Y. Liu are with Queen Mary University of London, London, UK (email:\{chao.zhang, w.yi, yuanwei.liu\}@qmul.ac.uk).}

\thanks{Z. Ding is with the School of Electrical and Electronic Engineering, The University of Manchester, Manchester, UK (e-mail: zhiguo.ding@manchester.ac.uk).}

\thanks{L. Song is with Department of Electronics, Peking University, Beijing, China (e-mail: lingyang.song@pku.edu.cn).}

\thanks{Part of this work has been submitted to IEEE Global Communications Conference (GLOBECOM), Madrid, Spain, December, 2021 \cite{conf}.}
}

\maketitle
\vspace{-1.5cm}
\begin{abstract}
Compared with the conventional reconfigurable intelligent surfaces (RIS), simultaneous transmitting and reflecting intelligent omini-surfaces (STAR-IOSs) are able to achieve $360^\circ$ coverage ``smart radio environments". By splitting the energy or altering the active number of STAR-IOS elements, STAR-IOSs provide high flexibility of successive interference cancellation (SIC) orders for non-orthogonal multiple access (NOMA) systems. Based on the aforementioned advantages, this paper investigates a STAR-IOS-aided downlink NOMA network with randomly deployed users. We first propose three tractable channel models for different application scenarios, namely the central limit model, the curve fitting model, and the M-fold convolution model. More specifically, the central limit model fits the scenarios with large-size STAR-IOSs while the curve fitting model is extended to evaluate multi-cell networks. However, these two models cannot obtain accurate diversity orders. Hence, we figure out the M-fold convolution model to derive accurate diversity orders. We consider three protocols for STAR-IOSs, namely, the energy splitting (ES) protocol, the time switching (TS) protocol, and the mode switching (MS) protocol. Based on the ES protocol, we derive closed-form analytical expressions of outage probabilities for the paired NOMA users by the central limit model and the curve fitting model. Based on three STAR-IOS protocols, we derive the diversity gains of NOMA users by the M-fold convolution model. The analytical results reveal that the diversity gain of NOMA users is equal to the active number of STAR-IOS elements. Numerical results indicate that 1) in high signal-to-noise ratio regions, the central limit model performs as an upper bound of the simulation results, while a lower bound is obtained by the curve fitting model; 2) the TS protocol has the best performance but requesting more time blocks than other protocols; 3) the ES protocol outperforms the MS protocol as the ES protocol has higher diversity gains.
\end{abstract}
\vspace{-0.7cm}


\vspace{-.7cm}
\section{Introduction}
\vspace{-.2cm}

As a promising technique of the six-generation cellular networks (6G) \cite{6GRIS1,6GRIS2,6GRIS3,RIS6Gmag}, conventional reconfigurable intelligent surfaces (RIS) are able to enhance the channel quality of current networks. By controlling the electromagnetic response of signals via the RISs, integrated signals are radiated and transmitted to the targeted direction. Hence, the reconfigurable environments are able to actively transfer and process information, which is named as ``smart radio environment (SRE)" in recent works \cite{yuanwei1,marco1}. Although RISs have extraordinary benefits for the 6G networks, one of the key challenges of RISs is that the conventional substrates of RISs are opaque items that may block the signals to the users behind the RISs, which results in a worse performance for the blocked users. To overcome this limitations, the recent development of meta-surfaces, namely the simultaneous transmitting and reflecting intelligent omini-surfaces (STAR-IOSs), allows signals to pass through substrates via refraction \cite{meta1,meta2,meta3,meta4,meta5}. Compared with the conventional RISs with half-space SREs, the STAR-IOSs realize a highly flexible full-space SRE \cite{xidong1,meta6}. Hence, independent reflection and refraction beamforming is able to be designed and integrated with high flexibility of STAR-IOS serving areas \cite{jiaqiletter,meta7}. In a word, STAR-IOSs bring the $360^{\circ}$ coverage of SREs into reality \cite{yuanwei1,STARadd1}.

To enhance the full-space coverage of SREs, three STAR-IOS protocols are proposed in recent works, namely the energy splitting (ES) protocol, the mode switching (MS) protocol, and the time switching (TS) protocol \cite{yuanwei1,xidong1}. Among the three protocols, the TS protocol exploits all STAR-IOS elements in different time blocks to separately reflect and transmit signals. The ES and MS protocols are capable of offering the flexibility of successive interference cancellation (SIC) orders for an advanced transmission scheme, namely non-orthogonal multiple access (NOMA). More specifically, the ES protocol manages the energy allocation among reflecting and transmitting links and the MS protocol activates different numbers of STAR-IOS elements for reflecting and transmitting links. After that, we are able to artificially differ the channel quality of the reflecting and transmitting links by different energy splitting coefficients via the ES protocol or by different numbers of active STAR-IOS elements via the MS protocol. With the ability of controlling the channel quality, when considering that a reflecting user and a transmitting user are paired in one NOMA cluster, STAR-IOSs help to adjust the SIC orders for satisfying different constraints, such as allowing the user with a high priority to obtain a high channel gain and maintaining a successful SIC process. Hence, the STAR-IOSs distinguish the SREs of NOMA users, which expands the applications of NOMA in the 6G networks efficiently \cite{RIS6G1,RIS6G2}. Hence, the evaluation of STAR-IOSs in NOMA systems are highly valuable to be investigated.

\vspace{-0.5cm}

\subsection{Related Works}
\vspace{-.2cm}

\subsubsection{Related Works for Conventional RIS-aided NOMA}

Recent research contributions have evaluated the conventional RIS-aided NOMA systems in several aspects. Several optimization methods are firstly proposed to cope with the integrated signals \cite{2beamforming1,2beamforming2,2beamforming3}. More specifically, the works propose joint passive beamforming designs for multi-cluster multiple-input-single-output (MISO) NOMA systems in \cite{2beamforming1}, for RIS enhanced massive NOMA systems in \cite{2beamforming2}, and for NOMA systems with user-ordering designs in \cite{2beamforming3}. With the aid of beamforming designs, the optimized physical channel models are evaluated by \cite{Passlossmodel,di2019reflection,chao}. The works in \cite{Passlossmodel,di2019reflection,chao} model the RIS-aided channels when the authors consider the RISs as linear materials. When considering the RISs as integrated antennas, channel models are proposed with performance analysis, such as channel models based on zero-forcing beamforming designs in \cite{tianwei1,ZF}. Under the presence of hardware impairment of RIS-aided networks, the physical layer performance analysis is investigated by deriving the outage probability and throughput expressions in \cite{hardware1}, and the security performance of a RIS-aided internet of things (IoT) NOMA network is also analyzed in \cite{hardware2}.

\subsubsection{Related Works for STAR-IOS-aided NOMA}

As a brand-new topic, only a few works have been investigated for the STAR-IOS-aided NOMA systems. The recent research focuses on the optimal beamforming designs of STAR-IOS networks based on power consumption minimization \cite{xidong1}, phase shift optimization \cite{ShuhangZhang2}, sum-rate maximization \cite{ShuhangZhang1}, and sum coverage range maximization \cite{xidong2}. Additionally, a joint design for STAR-IOS enhanced coordinated multi-point transmission (CoMP) NOMA systems is proposed by \cite{tianwei2}. As the STAR-IOSs improve the flexibility for downlink NOMA systems \cite{yuanwei1}, physical layer performance analysis is needed to derive valuable insights for finding out more optimization problems. However, obtaining tractable channel models is the main challenge for performance analysis of STAR-IOS-aided networks. Additionally, theoretical performance analysis for STAR-IOS-aided NOMA systems is still in their infancy.

\vspace{-.5cm}
\subsection{Motivation and Contributions}
\vspace{-.2cm}
To achieve the full-space coverage of SREs, we aim to first derive the tractable expressions for depicting physical channel models and to evaluate the physical layer performance. We additionally consider randomly deployed users to analyze the spatial effects in full-space SREs. Moreover, we utilize the ES, MS, and TS protocols to realize the application of STAR-IOSs. As the STAR-IOSs provide flexible SIC orders, we present three different channel models and investigate the outage performance of STAR-IOS-aided NOMA systems. Hence, the main contributions are summarized as follows:

 \begin{itemize}

 \item We derive three physical layer channel models for STAR-IOS-aided networks, i.e., the central limit model, the M-fold convolution model, and the curve fitting model. More specifically, we exploit the curve fitting model and the central limit model to analyze the outage performance. We conclude that the central limit model is utilized in the case with a large number of independent STAR-IOS elements, while the curve fitting model fits any scenarios but needs different curve fitting functions. As these two models cannot achieve accurate diversity orders, we additionally utilized the M-fold convolution model to evaluate asymptotic performance with diversity analysis.
  \item We exploit the curve fitting model and the central limit model to derive the closed-form expressions of the outage probability for NOMA users under the ES protocol. The analytical results indicate that the curve fitting model performs as a close lower bound of the simulation results while the central limit model is an upper bound.
  \item Under the ES, MS, and TS protocols, we derive the asymptotic outage probability expressions for NOMA users based on the M-fold convolution model. We then derive the diversity orders for the three protocols. The analytical results indicate that the diversity orders are equal to the active number of STAR-IOS elements.
  \item We verify our analytical results by Monte Carlo simulations. Numerical results demonstrate that: 1) STAR-IOSs enhance the outage performance of NOMA systems significantly and provide high flexibility for SIC orders; 2) the TS protocol has the best outage performance but it only serves one user in each time block; and 3) with two users served in the same resource block, the ES protocol outperforms the MS protocol as the diversity gains of the ES protocol is larger than that of the MS protocol.
\end{itemize}

\vspace{-.5cm}
\subsection{Organizations}
\vspace{-.2cm}
This paper is organized into the following sections. In Section II, we introduce a STAR-IOS-aided NOMA system model with randomly deployed users. In Section III, we derive three physical layer channel models, including the curve fitting model and the central limit model to evaluate approximated performance, and the M-fold convolution model to derive the diversity orders. In Section IV, we derive the closed-form outage probability expressions for NOMA users. In section V, we derive the asymptotic outage probability expressions with diversity analysis. We present numerical results in Section VI, followed by Section VII as a conclusion.

\vspace{-.5cm}
\section{System Model}
\vspace{-.2cm}

 \begin{figure*}[t]
 \vspace{-0.3cm}
\centering
\includegraphics[width= 4in]{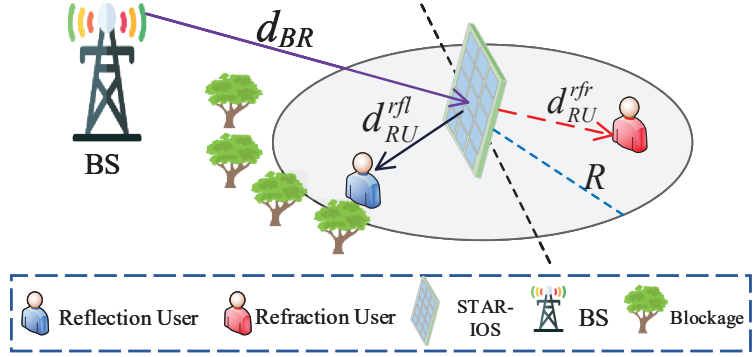}
\caption{Illustration of the considered STAR-IOS-aided downlink NOMA network: a fixed BS, a fixed STAR-IOS, and a pair of NOMA users (a transmitting user and a reflecting user) are considered. }
\label{fig_1}
\vspace{-0.5cm}
\end{figure*}

A STAR-IOS-aided downlink NOMA network is considered, which includes a fixed base station (BS), a fixed STAR-IOS, and randomly deployed users (reflecting users and transmitting users). We consider that a reflecting user and a transmitting user are paired in the same resource block with different power levels. We assume different NOMA pairs are allocated into orthogonal resource blocks, thereby inter-NOMA-cluster interference is ignored. We additionally assume the direct links for the two NOMA users are blocked. More specifically, for the reflecting user, the direct link is blocked by obstacles, such as trees or buildings. Thus, only the reflecting links via the STAR-IOS are included for the reflecting user. For the transmitting user, its location is behind the STAR-IOS substrates, thus the user is blocked. As the STAR-IOS is transparent for the signals of the transmitting user, the transmitting user still receives refracted signals with the aid of the STAR-IOS, which is the only approach for the transmitting user to receive signals. To sum up, the BS firstly transmits signals to the STAR-IOS, followed by the radiation signals to the reflecting user and the transmitting user.

\vspace{-.5cm}
\subsection{Theoretic Foundation of STAR-IOSs}
\vspace{-.2cm}
The main difference between STAR-IOSs and conventional RISs is that STAR-IOSs allow signals through themselves via refraction, which allows independent reflecting and refracting beamforming for the two half-spaces to achieve high flexibility \cite{jiaqiletter,yuanwei1,xidong1}. Hence, the serving area of STAR-IOSs is enhanced from a half circle (for the conventional RISs) to a whole circle area with the aid of simultaneous reflection and refraction. We define the reflecting and refracting coefficients as $R_m$ and $T_m$ for the $m^{th}$ STAR-IOS element, respectively. We consider that the phase shift coefficients are two independent variables, denoted as ${\phi _m^{rfl}}$ for the reflecting user and ${\phi _m^{rfr}}$ for the transmitting user. Additionally, we assume the STAR-IOS has $M$ elements satisfying $1 \le m\le M$. Hence, the reflected and refracted signals of the $m^{th}$ STAR-IOS element are expressed as ${R_m} = \sqrt {\beta _m^{rfl}} {e^{j\phi _m^{rfl}}}$ and $ {T_m} = \sqrt {\beta _m^{rfr}} {e^{j\phi _m^{rfr}}}$, respectively, where ${\phi _m^{rfl}},{\phi _m^{rfr}} \in [0,2\pi)$. Additionally, the ${\beta _m^{rfl}}$ is the energy coefficient for reflecting links and the ${\beta _m^{rfr}}$ is the energy coefficient for transmitting links.

Based on the theories in \cite{yuanwei1,xidong1}, we note that the STAR-IOS-aid networks have three practical protocols to be exploited, namely the ES, MS, and TS protocols. The detailed description and constraints are presented in the following.

\subsubsection{Energy Splitting Protocol of STAR-IOSs}

In terms of the ES protocol, we consider all the STAR-IOS elements ($M$ elements) simultaneously operate refraction and reflection modes, while the total radiation energy is split into two parts. When assuming the STAR-IOSs are passive with ignorable energy consumption, we present a constraint on the aforementioned coefficients as ${\left| {{R_m}} \right|^2} + {\left| {{T_m}} \right|^2} \le 1$. Hence, this protocol is mathematically presented as $\beta _m^{rfl} + \beta _m^{rfr} \le 1$ \cite{jiaqiletter}. As we expect the best utilization rate of STAR-IOS elements, we assume $\beta _m^{rfl} + \beta _m^{rfr} = 1$ in the following investigation \cite{xidong1}.

\subsubsection{Mode Switching Protocol of STAR-IOSs}

For the MS protocol, no STAR-IOS element will simultaneously reflect and refract signals. Instead, the STAR-IOS elements are partitioned into two groups. More specifically, the first group of STAR-IOS elements is exploited to fully reflect signals for reflecting links, while the other group of the STAR-IOS elements performs as the full refraction mode to be utilized in transmitting links. We assume that $M_{rfl}$ STAR-IOS elements are utilized for the reflecting links and $M_{rfr}$ STAR-IOS elements are exploited for the transmitting links. Hence, with the best utilization of STAR-IOS elements, we mathematically present the constraint of the MS protocol as $M_{rfl}+M_{rfr}=M$. For the $M_{rfl}$ elements, we have $\beta _m^{rfl}=1$ and $\beta _m^{rfr} =0$, while for the $M_{rfr}$ elements, we have $\beta _m^{rfl}=0$ and $\beta _m^{rfr} =1$.

\subsubsection{Time Switching Protocol of STAR-IOSs}

When it comes to the TS protocol, the $M$ elements are operated as the full refraction mode or the full reflection mode in different coherent time slots. For time slots with refraction modes, we have $\beta _m^{rfl}=0$ and $\beta _m^{rfr} =1$ and all the $M$ STAR-IOS elements perform as the full refraction mode. Additionally, we note $\beta _m^{rfl}=1$ and $\beta _m^{rfr} =0$ for time slots with the full reflection mode and all the STAR-IOS elements fully reflect signals. We define two time variation coefficients, namely $\lambda_{rfr}$ and $\lambda_{rfl}$ for the transmitting and reflecting links, respectively. Hence, we mathematically define the constraint as $\lambda_{rfr}+\lambda_{rfl}=1$ to present the percentage of time slot allocation.

\vspace{-.5cm}
\subsection{Deployment of Devices}
\vspace{-.2cm}
We consider a single-cell STAR-IOS-aided NOMA network. In this case, the BS is deployed at the center of the cell. Since the STAR-IOSs are always deployed at buildings facades, the positions of STAR-IOSs are fixed and known by us. We choose one of them to investigate the performance with the position denoted as $\mathbf{x_R}$. Then, we assume the STAR-IOSs are deployed on tall buildings. Thus, the links between the BS and the STAR-IOS elements are line of sight (LoS) links. For the users, the positions of the reflecting user and the transmitting user are expressed as $\mathbf{x_{rfl}}$ and $\mathbf{x_{rfr}}$, respectively. Note that direct links from the BS to the reflecting users are blocked by obstacles. Hence, we define the channel links as three types: 1) the link between the BS and the STAR-IOS as the $BR$ link with the distance $d_{BR}= \|\mathbf{x_R}\|$, 2) the link between the STAR-IOS and the reflecting user as the $RU_{rfl}$ link with the distance $d_{RU}^{rfl}= \|\mathbf{x_R}-\mathbf{x_{rfl}}\|$, and 3) the link from the STAR-IOS to the transmitting user as the $RU_{rfr}$ link with the distance $d_{RU}^{rfr}= \|\mathbf{x_R}-\mathbf{x_{rfr}}\|$.

We assume that users are uniformly distributed within the serving area of the STAR-IOS. Without loss of generality, we consider the serving area of the STAR-IOS is a circle with the radius $R$, denoted as $\mathbb{O}(0,R)$, where $\mathbb{O}(a,b)$ is an annulus with the inner radius $a$ and outer radius $b$. Additionally, this area is split into two parts: 1) the half ball facing the STAR-IOS as the reflecting area, namely $\mathbb{B}_{rfl}$, and 2) the rest half ball behind the STAR-IOS as the refracting area, namely $\mathbb{B}_{rfr}$. We randomly choose a user from $\mathbb{B}_{rfl}$ and a user from $\mathbb{B}_{rfr}$ as the NOMA pair. Hence, we evaluate the spatial effects of the chosen NOMA users. In this case, the probability density functions (PDFs) of $d_{RU}^{rfl}$ and $d_{RU}^{rfr}$ are expressed as
\begin{align}\label{rl}
{f_{d_{RU}^{rfl}}}\left( x \right)& = \frac{\partial }{{\partial x}}\int_0^\pi  {\int_0^x {\frac{{2r}}{{\pi {R^2}}}drd\theta } }  = \frac{{2x}}{{{R^2}}},\\
\label{rr}
{f_{d_{RU}^{rfl}}}\left( x \right) &= \frac{\partial }{{\partial x}}\int_\pi ^{2\pi } {\int_0^x {\frac{{2r}}{{\pi {R^2}}}drd\theta } }  = \frac{{2x}}{{{R^2}}}.
\end{align}

\vspace{-.5cm}
\subsection{Signal Model}
\vspace{-.2cm}
Based on the NOMA technique, the strong NOMA user in the NOMA pair accomplishes the SIC procedure. As the STAR-IOS is able to adjust the energy allocation coefficients $\beta _m^{rfl}$ and $\beta _m^{rfr}$ via the ES protocol, we allocate more energy for reflecting links. In practical scenarios, to ensure the links between the STAR-IOS and the users are LoS links, we assume the radius of STAR-IOS serving area $R$ is not large. Thus, the influence of path loss is not severe. Under this situation, we find a pair of energy allocation coefficients ($\beta _m^{rfl}$ and $\beta _m^{rfr}$) satisfying that the reflecting user is always kept as the strong user. Therefore, the reflecting user operates the SIC process. For the MS and TS protocol, we extend this assumption by allocating different numbers of active STAR-IOS elements or time blocks. Based on this assumption, the channel models are designed in the following.

\subsubsection{Small-scale Fading Model}
We assume that all the links in this model are Rician distribution. We denote the small scale fading of three types of links as $h_{BR,m}$ for BR links, $h_{RU,m}^{rfl}$ for $RU_{rfl}$ links, and $h_{RU,m}^{rfr}$ for $RU_{rfr}$ links for $\forall m \in \left\{ {1,2, \cdots ,M} \right\}$. Hence, the PDF for Rician distribution is expressed as
\begin{align}
{f_{{h_{BR,m}}}}\left( x \right) {=} {f_{h_{RU,m}^{rfl}}}\left( x \right) {=} {f_{h_{RU,m}^{rfr}}}\left( x \right){ =} \frac{{2\left( {1 + {k}} \right)}}{{\exp \left( {{k}} \right)}}x\exp \left[ { - \left( {1 + {k}} \right){x^2}} \right]{I_0}\left[ {2\sqrt {{k}\left( {1 + {k}} \right)} x} \right],
\end{align}
where $k$ is the coefficient of Rician distribution and $I_0(x)$ is the Bessel function. In this case, we assume that the mean values and variances of all the Rician channels are expressed as $\bar h = \sqrt {\frac{\pi }{{4\left( {1 + k} \right)}}} {}_1{F_1}\left( { - \frac{1}{2},1; - k} \right)$ and $\eta  = {\rm{1}} - \frac{\pi }{{4\left( {1 + k} \right)}}{\left[ {{}_1{F_1}\left( { - \frac{1}{2},1; - k} \right)} \right]^2}$, where ${}_1{F_1}\left( { \cdot , \cdot ; \cdot } \right)$ is the confluent hypergeometric function of the first kind.

We combine the $BR$ and $RU_{rfl}$ links as the reflecting link, namely $g_m^{rfl}$ for the $m^{th}$ STAR-IOS element. Additionally, we combine the $BR$ and $RU_{rfr}$ link as the transmitting link, namely $g_m^{rfr}$ for the $m^{th}$ STAR-IOS element. Based on the theoretic fundamental constraint of STAR-IOSs, we express the small scale fading model of the reflecting and transmitting links as
\begin{align}
\left| {g_m^{rfl}} \right| = \left| {{\bf{G}}_{{\bf{RU}}}^{{\bf{rfl}}}{{\bf{\Theta }}_{{\bf{rfl}}}}{{\bf{G}}_{{\bf{BR}}}}} \right|,\left| {g_m^{rfr}} \right| = \left| {{\bf{G}}_{{\bf{RU}}}^{{\bf{rfr}}}{{\bf{\Theta }}_{{\bf{rfr}}}}{{\bf{G}}_{{\bf{BR}}}}} \right|,
\end{align}
where ${{\bf{\Theta }}_{{\bf{rfl}}}} {=} diag\left[ {\sqrt {\beta _1^{rfl}} {e^{j\phi _1^{rfl}}},\sqrt {\beta _2^{rfl}} {e^{j\phi _2^{rfl}}}, \cdots ,\sqrt {\beta _M^{rfl}} {e^{j\phi _M^{rfl}}}} \right]$ is the diagonal matrix for reflecting links, ${{\bf{\Theta }}_{{\bf{rfr}}}} {=} diag\left[ {\sqrt {\beta _1^{rfr}} {e^{j\phi _1^{rfr}}},\sqrt {\beta _2^{rfr}} {e^{j\phi _2^{rfr}}}, \cdots ,\sqrt {\beta _M^{rfl}} {e^{j\phi _M^{rfl}}}} \right]$ is the diagonal matrix for transmitting links, ${\bf{G}}_{{\bf{RU}}}^{{\bf{rfl}}}  {=} {\left[ {h_{RU,1}^{rfl},h_{RU,2}^{rfl}, \cdots ,h_{RU,M}^{rfl}} \right]^T}$, ${\bf{G}}_{{\bf{RU}}}^{{\bf{rfr}}} {=} {\left[ {h_{RU,1}^{rfr},h_{RU,2}^{rfr}, \cdots ,h_{RU,M}^{rfr}} \right]^T}$, and ${{\bf{G}}_{{\bf{BR}}}} = \left[ {{h_{BR,1}},{h_{BR,2}}, \cdots ,{h_{BR,M}}} \right]$.
\subsubsection{STAR-IOS-aided Path Loss Model}

We define the path loss model of the three links via conventional wireless communication models. Hence, we respectively express the path loss expressions for $BR$, $RU_{rfl}$, and $RU_{rfr}$ links as
 \begin{align}
&{{\cal P}_{BR}}({{\bf{x}}_R}) = {C_{BR}}{\left\| {{{\bf{x}}_\mathbf{R}}} \right\|^{ - {\alpha _t}}} = {C_{BR}}d_{BR}^{ - {\alpha _t}},\\
&{\cal P}_{RU}^{rfl}({{\bf{x}}_R},{\bf{x}}_\mathbf{{RU}}^\mathbf{{rfl}}) = C_{RU}^{rfl}{\left\| {{{\bf{x}}_R} - {\bf{x}}_\mathbf{{RU}}^\mathbf{{rfl}}} \right\|^{ - {\alpha _t}}} = C_{RU}^{rfl}{\left( {d_{RU}^{rfl}} \right)^{ - {\alpha _t}}},\\
&{\cal P}_{RU}^{rfr}({{\bf{x}}_\mathbf{R}},{\bf{x}}_\mathbf{{RU}}^\mathbf{{rfr}}) = C_{RU}^{rfr}{\left\| {{{\bf{x}}_R} - {\bf{x}}_\mathbf{{RU}}^\mathbf{{rfr}}} \right\|^{ - {\alpha _t}}} = C_{RU}^{rfr}{\left( {d_{RU}^{rfr}} \right)^{ - {\alpha _t}}},
 \end{align}
where the ${\cal P}$ expresses the path loss, $ \{C_{BR},C_{RU}^{rfl},C_{RU}^{rfr}\}={\left( {\frac{c}{{4\pi {f_c}}}} \right)^2}$ are reference-distance based intercepts for different links and the reference distance $d_0 = 1$ m in this work, in which $c=3\times10^8$ m/s is the speed of light and $f_c$ is the used carrier frequency. Additionally, the $\alpha_t$ is the path loss exponent for users.

\subsubsection{Signal-to-Interference-and-Noise Ratio (SINR)}

To ensure the strong user (the reflecting user) having the SIC process, we allocate more transmit power to the week user (the transmitting user) by the BS. Hence, the SINR of the SIC process for the reflecting user is given by
\begin{align}
{\gamma _{{\rm{SIC}}}} = \frac{{{a_{rfr}}{P_t}{{\cal P}_{BR}}({{\bf{x}}_R}){\cal P}_{RU}^{rfl}({{\bf{x}}_R},{\bf{x}}_{RU}^{rfl}){{\left| {g_m^{rfl}} \right|}^2}}}{{{a_{rfl}}{P_t}{{\cal P}_{BR}}({{\bf{x}}_R}){\cal P}_{RU}^{rfl}({{\bf{x}}_R},{\bf{x}}_{RU}^{rfl}){{\left| {g_m^{rfl}} \right|}^2} + {\sigma ^2}}},
 \end{align}
where $P_t$ is the transmit power of the BS, $\sigma^2$ is the variance of additive white Gaussian noise (AWGN), and ${{a_{rfr}}}$ and ${{a_{rfl}}}$ are transmit power allocation coefficient satisfying ${{a_{rfr}}}+{{a_{rfl}}}=1$ and ${{a_{rfl}}}<{{a_{rfr}}}$.

With the aid of SIC, the reflecting user removes the messages of the transmitting user. Then, the reflecting user decodes its required messages. Hence, the signal-to-noise ratio (SNR) of the reflecting user is presented as
\begin{align}
{\gamma _{{\rm{rfl}}}} = \frac{{{a_{rfl}}{P_t}{{\cal P}_{BR}}({{\bf{x}}_R}){\cal P}_{RU}^{rfl}({{\bf{x}}_R},{\bf{x}}_{RU}^{rfl}){{\left| {g_m^{rfl}} \right|}^2}}}{{{\sigma ^2}}}.
\end{align}

When we consider the other NOMA user's messages as interference, the transmitting user directly decodes its signal. Hence, the SINR of the transmitting user is expressed as
\begin{align}
{\gamma _{{\rm{rfr}}}} = \frac{{{a_{rfr}}{P_t}{{\cal P}_{BR}}({{\bf{x}}_R}){\cal P}_{RU}^{rfr}({{\bf{x}}_R},{\bf{x}}_{RU}^{rfr}){{\left| {g_m^{rfr}} \right|}^2}}}{{{a_{rfl}}{P_t}{{\cal P}_{BR}}({{\bf{x}}_R}){\cal P}_{RU}^{rfr}({{\bf{x}}_R},{\bf{x}}_{RU}^{rfr}){{\left| {g_m^{rfr}} \right|}^2} + {\sigma ^2}}}.
 \end{align}

Based on the aforementioned SINR expressions, our first target is to derive the STAR-IOS-aided channel models in the following section.
\vspace{-.3cm}
\section{STAR-IOS-aided Channel Model Approximation}

\vspace{-.1cm}

As the exact channel models of STAR-IOS-aided networks are complex, it is important to derive approximated mathematical channel models that are tractable for performance analysis. Hence, we present three approximated models for different application scenarios, namely the central limit model, the M-fold convolution model, and the curve fitting model. More specifically, the M-fold convolution model is used to derive diversity orders. The central limit model is suitable for large STAR-IOSs with a large number of elements. For the curve fitting model, it fits all scenarios by adjusting the curve fitting functions and parameters but exploring a suitable curve fitting function is a challenge. Detailed derivations and discussions are expressed in the following subsections.

\vspace{-.5cm}
\subsection{Central Limit Model}
\vspace{-.2cm}
When we assume the channel gains of all the elements of STAR-IOSs are irrelevant, the channel model of STAR-IOS-aided networks is expressed as the summation of different variables. Hence, the central limit theorem is an appropriate mathematical tool to derive the approximated channel model. Although it has the constraint that the STAR-IOS elements are large enough, the central limit model is one of the most popular models in recent works because of its great tractability on derivations. Hence, under the case with quantities of uncorrelated channels passing by different STAR-IOS elements, we exploit the central limit model to investigate the channel performance.

\begin{lemma}\label{channel1}
 We assume that the quantity of STAR-IOS elements $M$ is large and the channels for different STAR-IOS elements are independent. For the ES protocol and with the aid of the central limit theorem, the PDF and cumulative distribution function (CDF) of the central limit model are derived as
\begin{align}
{f_{{{\left| {g_m^{rf}} \right|}^2}}}\left( y \right) &= \frac{1}{{{\rm{2}}\sqrt {2\pi \eta _{eq}^{rf}} }}\frac{1}{{\sqrt y }}\left( {\exp \left( { - \frac{{{{\left( {\sqrt y  - \bar h_{eq}^{rf}} \right)}^2}}}{{2\eta _{eq}^{rf}}}} \right){\rm{ + }}\exp \left( { - \frac{{{{\left( {\sqrt y {\rm{ + }}\bar h_{eq}^{rf}} \right)}^2}}}{{2\eta _{eq}^{rf}}}} \right)} \right),\\
{F_{{{\left| {g_m^{rf}} \right|}^2}}}\left( y \right) &= \frac{1}{{\rm{2}}}\left( {\erf\left( {\frac{{\bar h_{eq}^{rf} + \sqrt x }}{{\sqrt {2\eta _{eq}^{rf}} }}} \right) - \erf\left( {\frac{{\bar h_{eq}^{rf} - \sqrt x }}{{\sqrt {2\eta _{eq}^{rf}} }}} \right)} \right),
\end{align}
where ${\bar h_{eq}^{rf}}$ is the mean value of $|{g_m^{rf}}|$ with $rf \in \left\{ {rfr,rfl} \right\}$ representing the transmitting links and the reflecting links, respectively. The $\eta _{eq}^{rf}$ is the variance of $|{g_m^{rf}}|$. Based on the properties of the expectation and the variance for independent variables, we derive $\bar h_{eq}^{rf} =\sqrt {{\beta _{rf}}} M{{\bar h}^2}$ and $\eta _{eq}^{rf} = {\beta _{rf}}M\left( {2{{\bar h}^2}\eta  + {\eta ^2}} \right)$. Additionally, the function $\erf (\cdot)$ is the error function.
\begin{IEEEproof}
See Appendix~A.
\end{IEEEproof}
\end{lemma}

\begin{remark}
For the MS and TS protocols, the central limit channel model expressions are almost the same but the coefficients are different. More specifically, for the MS protocol, we have $\bar h_{eq,MS}^{rf} = {M_{rf}}{{\bar h}^2}$ and $\eta _{eq,MS}^{rf} = {M_{rf}}\left( {2{{\bar h}^2}\eta  + {\eta ^2}} \right)$ with $rf \in \{rfr, rfl\}$ for transmitting links and reflecting links, respectively. For the TS model, the channel model coefficients are expressed as $\bar h_{eq,TS}^{rf} = M{{\bar h}^2}$ and $\eta _{eq,TS}^{rf} = M\left( {2{{\bar h}^2}\eta  + {\eta ^2}} \right)$.
\end{remark}

\vspace{-.5cm}
\subsection{M-Fold Convolution Model}
\vspace{-.2cm}

When analyzing the performance of a system, we always consider the diversity orders to evaluate the performance in high SNR regions. Although the central limit model performs as a well-matched channel model with closed-form expressions, we cannot obtain the accurate diversity orders as the high SNR region is not perfectly matching. Hence, exploiting the Laplace transform, we achieve the accurate diversity orders by the M-fold convolution model.

\begin{lemma}\label{channel2}
To investigate the diversity orders, we utilize the M-fold convolution method to derive the STAR-IOS-aided channel model. We denote the Rician coefficient of the BR links as $k_1$ and that of the RU links as $k_2$. We utilize the ES protocol in this theorem. In high SNR regions, we derive the approximated PDF and CDF expressions as
\begin{align}
f_{{{\left| {g_m^{rf}} \right|}^2},{ES}}^{0 + }\left( x \right) &= \frac{{{{\left[ {\sigma \left( {0,0} \right)} \right]}^M}{x^{M - 1}}}}{{2{{\left( {{\beta _{rf}}} \right)}^M}\left( {2M - 1} \right)!}},\\
F_{{{\left| {g_m^{rf}} \right|}^2},{ES}}^{0 + }\left( x \right) &= \frac{{{{\left[ {\sigma \left( {0,0} \right)} \right]}^M}}}{{2{{\left( {{\beta _{rf}}} \right)}^M}M\left( {2M - 1} \right)!}}{x^M},
\end{align}
where $\sigma \left( {t,n} \right) = \frac{{{4^{t - n + 1}}\sqrt \pi  k_1^tk_2^t{{\left[ {\left( {1 + {k_{\rm{1}}}} \right)\left( {1 + {k_{\rm{2}}}} \right)} \right]}^{t + 1}}}}{{{{\left( {t!} \right)}^2}{{\left( {n!} \right)}^2}\exp \left( {{k_{\rm{1}}} + {k_{\rm{2}}}} \right)}}\Gamma \binom{2n + 2,2t + 2}{t + n + \frac{5}{2}}{}_2{F_1}\left( {2t + 2,t - n + \frac{1}{2};t + n + \frac{5}{2};1} \right)$, and ${}_2{F_1}\left( { \cdot , \cdot ; \cdot ; \cdot } \right)$ is the ordinary hypergeometric function. Additionally, the $k_1$ and $k_2$ are the Rician coefficients for the channels of BR links and RU links, respectively. If we consider all the channel links with the same Rician distribution with parameter $k$, the parameter is further derived as $\sigma \left( {t,n} \right) = \frac{{{4^{t - n + 1}}\sqrt \pi  {k^{{\rm{2t}}}}{{\left( {1 + k} \right)}^{2(t + 1)}}}}{{{{\left( {t!} \right)}^2}{{\left( {n!} \right)}^2}\exp \left( {2k} \right)}}\Gamma \binom{2n + 2,2t + 2}{t + n + \frac{5}{2}}{}_2{F_1}\left( {2t + 2,t - n + \frac{1}{2};t + n + \frac{5}{2};1} \right)$.
\begin{IEEEproof}
See Appendix~B.
\end{IEEEproof}
\end{lemma}

\begin{corollary}
For the MS protocol, the M-fold convolution channel model is derived as
\begin{align}
f_{{{\left| {g_m^{rf}} \right|}^2},{MS}}^{0 + }\left( x \right) &= \frac{{{{\left[ {\sigma \left( {0,0} \right)} \right]}^{M_{rf}}}{x^{M_{rf} - 1}}}}{{2\left( {2M_{rf} - 1} \right)!}},\\
F_{{{\left| {g_m^{rf}} \right|}^2},{MS}}^{0 + }\left( x \right) &= \frac{{{{\left[ {\sigma \left( {0,0} \right)} \right]}^{M_{rf}}}}}{{2M_{rf}\left( {2M_{rf} - 1} \right)!}}{x^{M_{rf}}},
\end{align}
where $rf \in \{rfr, rfl\}$ is presented as transmitting links and reflecting links, respectively. We note that $M_{rfr}+M_{rfl}=M$ to fully utilize the STAR-IOS elements.

Additionally, for the TS protocol, the M-fold convolution channel model is derived as
\begin{align}
f_{{{\left| {g_m^{rf}} \right|}^2},{TS}}^{0 + }\left( x \right) &= \frac{{{{\left[ {\sigma \left( {0,0} \right)} \right]}^M}{x^{M - 1}}}}{{2\left( {2M - 1} \right)!}},\\
F_{{{\left| {g_m^{rf}} \right|}^2},{TS}}^{0 + }\left( x \right) &= \frac{{{{\left[ {\sigma \left( {0,0} \right)} \right]}^M}}}{{2M\left( {2M - 1} \right)!}}{x^M}.
\end{align}
\begin{IEEEproof}
The proof is similar to \textbf{Lemma \ref{channel2}}.
\end{IEEEproof}
\end{corollary}

\vspace{-.7cm}
\subsection{Curve Fitting Model}
\vspace{-.2cm}
The third approximated channel model is denoted as the curve fitting model. We exploit the Matlab curve fitting tool to mimic the channel model as an extant distribution. More specifically, independent channels of different STAR-IOS elements are considered, thus we approximately mimic the channel model as a Gamma distribution. In a word, the curve fitting model has wider application scenarios compared to the aforementioned two channel models, while it has new challenges.

Compared to the M-fold convolution, the accurate diversity gains cannot be obtained by the curve fitting model as the curve fitting model does not match the high SNR regions well. Hence, the M-fold convolution model is still irreplaceable for diversity analysis.

Compared to the central limit model, the curve fitting model has pros and cons as follows. The advantage of the curve fitting model is that it suits more scenarios and tractable for further derivation, i.e., the STAR-IOS-aided networks with few elements and multi-cell scenarios\footnote{The central limit model cannot match STAR-IOS-aided networks with few elements, thereby we need a channel model to be exploited with few STAR-IOS elements. Additionally, we expect that the channel models have exponential functions, which are tractable for multi-cell scenarios as we always use the Laplace transform to calculate the interference.}. Additionally, it even fits the scenarios when the STAR-IOS elements are influenced by each other with different curve fitting functions. However, the disadvantage is that the curve fitting model does not include detailed mathematical derivations, and exploring an accurate distribution may be challenging in some specific cases. Moreover, it is hard to ensure that we are able to find the best curve fitting function to make it more accurate than the central limit model. Hence, we should select different channel models based on specific scenarios but not advocate exploiting anyone.

\begin{lemma}\label{channel3}
Utilizing the curve fitting tool, it indicates that the combined channel ${{{\left| {g_m^{rf}} \right|}^2}}$ is able to be simulated as the Gamma distribution with the element $\alpha$ and $\beta$. Under the ES protocol, the PDF and CDF for the curve fitting model are expressed as
\begin{align}\label{channel31}
{f_{{{\left| {g_m^{rf}} \right|}^2}}}\left( x \right)& = \frac{{{x^{\alpha  - 1}}}}{{\Gamma \left( \alpha  \right){{\left( {{\beta _{rf}}\beta } \right)}^\alpha }}}\exp \left( { - \frac{x}{{{\beta _{rf}}\beta }}} \right),\\
\label{channel32}
{F_{{{\left| {g_m^{rf}} \right|}^2}}}\left( x \right) &= \frac{{\gamma \left( {\alpha ,\frac{x}{{{\beta _{rf}}\beta }}} \right)}}{{\Gamma \left( \alpha  \right)}},
\end{align}
where ${\gamma \left( {\cdot ,\cdot } \right)}$ is the incomplete Gamma function and $\Gamma(\cdot)$ is the Gamma function. Based on the mathematical tool, we observe that $\alpha=M$ and $\beta<M$, e.g., $\alpha=30$ and $\beta=22.46$ when the number of STAR-IOS elements $M=30$. The detailed values of the coefficient $\beta$ should be further calculated by the Matlab curve fitting tools.
\begin{IEEEproof}
Because of ${\left| {g_m^{rf}} \right|^2} = {\beta _{rf}}{\left( {\sum\limits_{m = 1}^M {h_{RU,m}^{rf}{h_{BR,m}}} } \right)^2}$ with $rf \in \left\{ {rfl,rfr} \right\}$ representing the reflecting and transmitting links, we first utilize the curve fitting tool to mimic the variable of $\frac{|g_m^{rf}|^2}{\beta _{rf}} $ as a Gamma distribution. Hence, we express the PDF of $\frac{|g_m^{rf}|^2}{\beta _{rf}}$ as
\begin{align}\label{t3}
{f_{{{{{\left| {g_m^{rf}} \right|}^2}} \mathord{\left/
 {\vphantom {{{{\left| {g_m^{rfl}} \right|}^2}} {{\beta _{rf}}}}} \right.
 \kern-\nulldelimiterspace} {{\beta _{rf}}}}}}\left( x \right) = \frac{{{x^{\alpha  - 1}}}}{{\Gamma \left( \alpha  \right){\beta ^\alpha }}}\exp \left( { - \frac{x}{\beta }} \right).
\end{align}

Based on \eqref{t3}, we then derive the final PDF and CDF of ${{{\left| {g_m^{rf}} \right|}^2}}$ as \eqref{channel31} and \eqref{channel32} in this theorem and this proof is end.
\end{IEEEproof}
\end{lemma}

\begin{remark}
For the MS and TS protocols, the curve fitting channel model expressions are the same as the expressions for the ES protocol, while the coefficients should be changed. In detail, for the MS protocol, we have $\alpha= M_{rfr}$ and $\alpha=M_{rfl}$ for the transmitting links and reflecting links, respectively. Additionally, the coefficient $\beta_{rf}=1$ as no energy allocation is included. For the TS protocol, the channel model coefficients are expressed as $\alpha = M$ and $\beta_{rf}=1$. The coefficient $\beta$ under the MS and TS protocols should be further determined by the Matlab curve fitting tools.
\end{remark}

\vspace{-.5cm}
\subsection{Comparison}
\vspace{-.2cm}

In this section, we compare the accuracy and complexity of the three channel models. For diversity analysis, since the curve fitting model and the central limit model have changed the channel distributions and do not match the high SNR regions well \cite{tianwei1}, the M-fold convolution model has more accurate diversity orders\footnote{As the central limit model and the curve fitting model have changed the distributions for the STAR-IOS channels, they cannot fit the high SNR regions well. The M-fold convolution derives the original distribution of the IOS channels, thus it has accurate diversity orders.}. Additionally, for asymptotic analysis, as the M-fold convolution model exploits the Taylor series to derive the asymptotic expressions, the complexity of the M-fold convolution model is the lowest among three models.

\begin{figure}[!htb]
\vspace{-0.4cm}
\centering
\includegraphics[width= 3.2in]{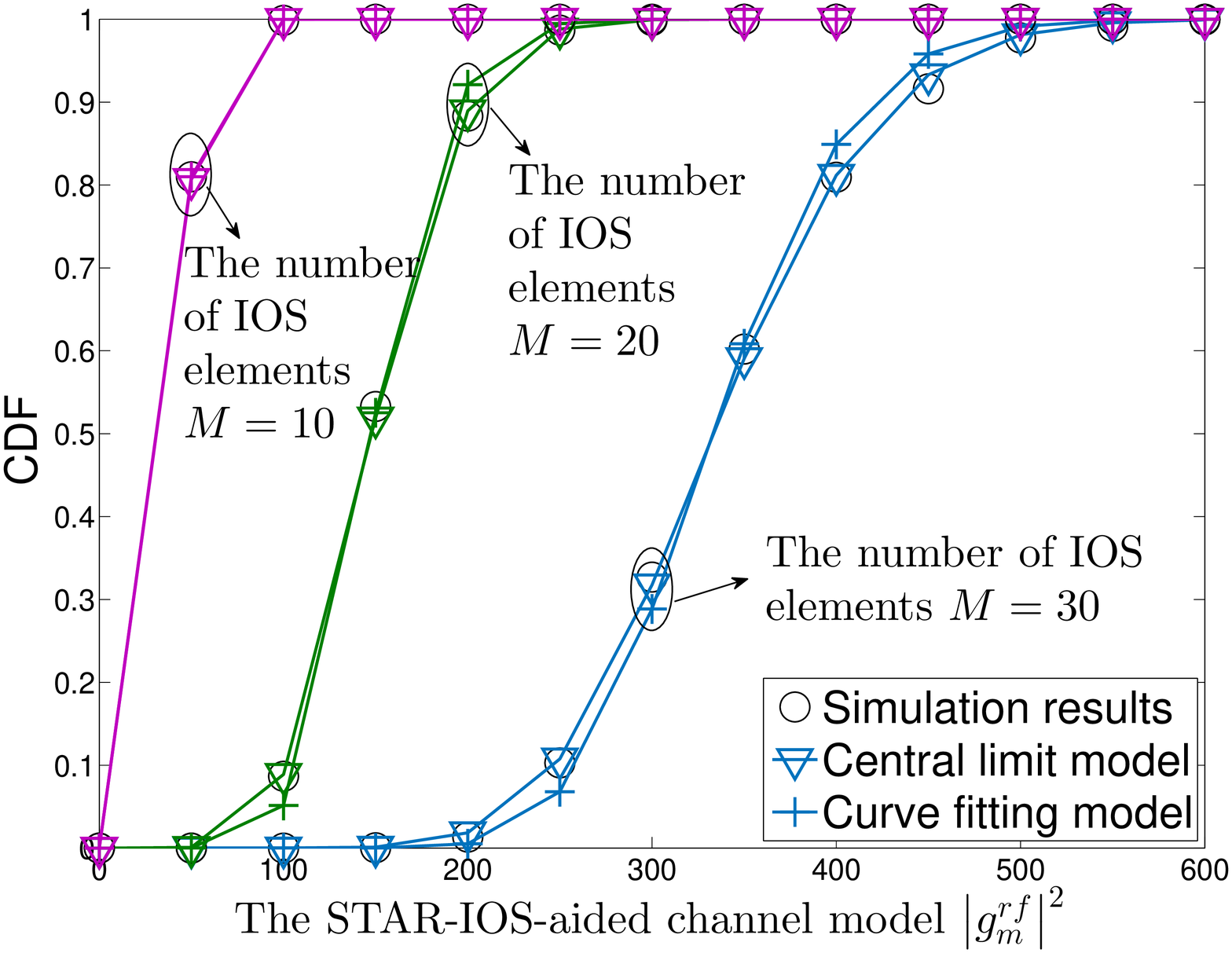}
\vspace{-0.5cm}
\caption{A CDF comparison between the curve fitting model and the central limit model with various numbers of STAR-IOS elements $M=\{10,20,30\}$.}
\label{figure2}
\vspace{-1cm}
\end{figure}

Then, we compare the CDFs of the central limit model and the curve fitting model. For the accuracy, Fig. \ref{figure2} demonstrates that the central limit model is more accurate than the curve fitting model. Thus, the central limit model has more accurate outage performance than the curve fitting model. For the complexity, as the curve fitting model mimics the STAR-IOS channel as a Gamma distribution, we are able to exploit current methods to evaluate the outage performance, especially for multi-cell scenarios. Hence, the complexity of the curve fitting model on performance analysis is lower than the central limit model.

To sum up, we exploit the central limit model to obtain best accuracy of outage performance; we harness the curve fitting model to obtain tractable derivations (especially for multi-cell scenarios); and we utilize the M-fold convolution model to obtain accurate diversity gains.

\vspace{-.3cm}
\section{Outage Performance Analysis}
\vspace{-.1cm}

In this section, we investigate the outage performance of the STAR-IOS-aided downlink NOMA network. More specifically, we consider exploiting the central limit model and the curve fitting model to calculate the approximated expressions of the outage probability for the reflecting and transmitting users.

In the following, we utilize the ES protocol to calculate the approximated outage probability via the central limit model and the curve fitting model. The outage performance of the MS protocol and the TS protocol is investigated with the same approaches of \textbf{Theorem \ref{clml}} to \textbf{Theorem \ref{cfmr}}, while the details are omitted due to space limitations.

Based on the NOMA technique, we first express the definition of the outage probabilities of the reflecting and transmitting users as
\begin{align}\label{Poutdef_rfl}
P_{out,rfl}^{}\left( x \right) &= 1 - \Pr \left\{ {{\gamma _{{\rm{SIC}}}} > \gamma _{th}^{SIC},{\gamma _{rfl}} > \gamma _{th}^{out}} \right\},\\
\label{Poutdef_rfr}
P_{out,rfr}^{}\left( x \right) &= \Pr \left\{ {{\gamma _{rfr}} < \gamma _{th}^{out}} \right\},
\end{align}
where $\gamma _{th}^{SIC}$ is the threshold for the SIC process, $\gamma _{th}^{out}$ is the outage threshold. We then calculate the closed-form outage probability expressions based on the three aforementioned models as follows.

\vspace{-0.5cm}
\subsection{Central Limit Model}
\vspace{-0.2cm}
As we have calculated the channel model by \textbf{Lemma \ref{channel1}}, we first utilize the central limit model to calculate the approximated outage probability expressions of the reflecting and transmitting users. Note that we have defined the power allocation coefficients as ${a_{rfr}}$ and ${a_{rfl}}$ for the transmitting user and reflecting user, respectively. Based on the outage probability definitions, we obtain that $P_{out,rfl}^{}\left( x \right)=1$ when ${a_{rfr}} < \gamma _{th}^{SIC}{a_{rfl}}$ and $P_{out,rfr}^{}\left( x \right) =1$ when ${a_{rfr}} < \gamma _{th} {a_{rfl}}$.

As we split more energy by the STAR-IOS elements to the reflecting user, it performs as the strong user. With the SIC process, we evaluate the outage probability of the reflecting user as the \textbf{Theorem \ref{clml}}. Without the SIC process, the outage probability expression of the transmitting user is derived as \textbf{Theorem \ref{clmr}}.

\vspace{-0.2cm}
\begin{theorem}\label{clml}
We note that $P_{out,rfl}^{}\left( x \right)=1$ when ${a_{rfr}} < \gamma _{th}^{SIC}{a_{rfl}}$. Under the case as ${a_{rfr}} > \gamma _{th}^{SIC}{a_{rfl}}$, we derive the closed-form outage probability expression of the reflecting user under the ES protocol as
\begin{align}
\vspace{-0.5cm}
P_{out,rfl}^{cl} = &\sum\limits_{n = 0}^\infty  {\frac{{4{{\left( { - 1} \right)}^n}}}{{n!\sqrt \pi  \left( {2n + 1} \right){{\left( {2\eta _{eq}^{rfl}} \right)}^{\frac{{2n + 1}}{2}}}}}}\notag\\
&\times \sum\limits_{r = \{ 1,3, \cdots, 2n + 1\} }^{2n + 1} {\binom{2n+1}{r}} \frac{{{R^{\frac{{{\alpha _t}r}}{2}}}{{\left( {\bar h_{eq}^{rfl}} \right)}^{2n + 1 - r}}}}{{\frac{{{\alpha _t}r}}{2} + 2}}{\left( {\frac{{{\Upsilon _{\max }}d_{BR}^{{\alpha _t}}}}{{{P_t}{C_{BR}}C_{RU}^{rfl}}}} \right)^{\frac{r}{2}}},
\end{align}
where ${\Upsilon _{\max }} = \max \left\{ {\frac{{\gamma _{th}^{SIC}{\sigma ^2}}}{{{a_{rfr}} - \gamma _{th}^{SIC}{a_{rfl}}}},\frac{{\gamma _{th}^{out}{\sigma ^2}}}{{{a_{rfl}}}}} \right\}$, ${\gamma _{th}^{out}}$ is the outage threshold and $\binom{n}{r}=\frac{n!}{r!(n-r)!}$.
\begin{IEEEproof}
See Appendix~C.
\end{IEEEproof}
\end{theorem}

\vspace{-0.2cm}
\begin{theorem}\label{clmr}
We note that $P_{out,rfr}^{}\left( x \right)=1$ when ${a_{rfr}} < \gamma _{th}{a_{rfl}}$. Under the case as ${a_{rfr}} > \gamma _{th}{a_{rfl}}$ and with the aid of the ES protocol, the closed-form outage probability expression of the transmitting user is derived as
\begin{align}
\vspace{-0.5cm}
P_{out,rfr}^{cl} = &\sum\limits_{n = 0}^\infty  {\frac{{4{{\left( { - 1} \right)}^n}}}{{n!\sqrt \pi  \left( {2n + 1} \right){{\left( {2\eta _{eq}^{rfr}} \right)}^{\frac{{2n + 1}}{2}}}}}} \sum\limits_{r = \{ 1,3, \cdots, 2n + 1\} }^{2n + 1} {\binom{2n+1}{r}} \notag\\
&\times \frac{{{R^{\frac{{{\alpha _t}r}}{2}}}{{\left( {\bar h_{eq}^{rfr}} \right)}^{2n + 1 - r}}}}{{\frac{{{\alpha _t}r}}{2} + 2}}{\left( {\frac{{{\Upsilon _2}d_{BR}^{{\alpha _t}}}}{{{P_t}{C_{BR}}C_{RU}^{rfr}}}} \right)^{\frac{r}{2}}},
\end{align}
where $\Upsilon _2=\frac{{\gamma _{th}^{out}{\sigma ^2}}}{{{a_{rfr}} - \gamma _{th}^{out}{a_{rfl}}}}$.
\begin{IEEEproof}
With the aid of \eqref{rl} and the Taylor series of the error function, following the process of \textbf{Theorem \ref{clml}}, this outage probability of the transmitting user is derived as
\begin{align}
P_{out,rfr}^{cl}\left( x \right)=&\frac{{\rm{2}}}{{\sqrt \pi  {R^2}}}\int_0^R {x\left( {\sum\limits_{n = 0}^\infty  {\frac{{{{\left( { - 1} \right)}^n}}}{{n!\left( {2n + 1} \right)}}} {{\left( {\frac{{\bar h_{eq}^{rfr} + \sqrt {\frac{{{\Upsilon _2}d_{BR}^{{\alpha _t}}{x^{{\alpha _t}}}}}{{{P_t}{C_{BR}}C_{RU}^{rfr}}}} }}{{\sqrt {2\eta _{eq}^{rfr}} }}} \right)}^{2n + 1}}} \right.} \notag\\
&\left. { - \frac{2}{{\sqrt \pi  }}\sum\limits_{n = 0}^\infty  {\frac{{{{\left( { - 1} \right)}^n}}}{{n!\left( {2n + 1} \right)}}} {{\left( {\frac{{\bar h_{eq}^{rfr} - \sqrt {\frac{{{\Upsilon _2}d_{BR}^{{\alpha _t}}{x^{{\alpha _t}}}}}{{{P_t}{C_{BR}}C_{RU}^{rfr}}}} }}{{\sqrt {2\eta _{eq}^{rfr}} }}} \right)}^{2n + 1}}} \right)dx.
\end{align}
and this theorem is proved based on the binomial theorem.
\end{IEEEproof}
\end{theorem}

\vspace{-0.7cm}
\subsection{Curve Fitting Model}
\vspace{-0.2cm}

As the central limit theorem only fits the scenario when the STAR-IOS is large with plenties of STAR-IOS elements. For those STAR-IOSs with few elements, the central limit model is not accurate. Hence, we consider the curve fitting model to cope with this problem. Additionally, the curve fitting model can be exploited into multi-cell networks since it is able to mimic the multi-cell networks as distributions with an exponential function to calculate the Laplace transform of interference.

We first note that $P_{out,rfl}^{}\left( x \right)=1$ when ${a_{rfr}} < \gamma _{th}^{SIC}{a_{rfl}}$ and $P_{out,rfr}^{}\left( x \right)=1$ when ${a_{rfr}} < \gamma _{th}{a_{rfl}}$. Hence, in the following theorems, we calculate the outage probability of the reflecting and transmitting users, respectively, when the outage probability is not constantly equal to one. We focus on the reflecting user in \textbf{Theorem \ref{cfml}}. Then, we utilize the same channel model to investigate the outage performance of the transmitting user. Hence, we derive the outage probability of the transmitting user in \textbf{Theorem \ref{cfmr}}.
\vspace{-0.2cm}
\begin{theorem}\label{cfml}
We consider the scenario that all the channels through different STAR-IOS elements are independent. Based on the curve fitting model with the ES protocol, we modeled the STAR-IOS channel as a Gamma distribution. Under the case as ${a_{rfr}} > \gamma _{th}^{SIC}{a_{rfl}}$, we derive the closed-form outage probability expression of the reflecting user as
\begin{align}
P_{out,rfl}^{cf} = \frac{{\rm{2}}}{{\Gamma \left( \alpha  \right)}}\sum\limits_{n = 0}^\infty  {\frac{{{{\left( { - 1} \right)}^n}{R^{{\alpha _t}\left( {\alpha  + n} \right)}}}}{{n!\left( {\alpha  + n} \right)\left[ {{\alpha _t}\left( {\alpha  + n} \right) + 2} \right]}}{{\left( {\frac{{{\Upsilon _{\max }}d_{BR}^{{\alpha _t}}}}{{{P_t}{C_{BR}}C_{RU}^{rfl}{\beta _{rfl}}\beta }}} \right)}^{\alpha  + n}}} .
\end{align}
\begin{IEEEproof}
See Appendix~D.
\end{IEEEproof}
\end{theorem}

\begin{theorem}\label{cfmr}
We consider the same scenario in \textbf{Theorem \ref{cfml}}, that is, the independent channels among the STAR-IOS elements. Additionally, we consider the case ${a_{rfr}} > \gamma _{th}{a_{rfl}}$. Following the aforementioned constraints with the ES protocol, we derive the closed-form outage probability expression of the transmitting user as
\begin{align}
P_{out,rfr}^{cf} = \frac{{\rm{2}}}{{\Gamma \left( \alpha  \right){R^{\rm{2}}}}}\sum\limits_{n = 0}^\infty  {\frac{{{{\left( { - 1} \right)}^n}}}{{n!\left( {\alpha  + n} \right)}}} {\left( {\frac{{{\Upsilon _{\rm{2}}}d_{BR}^{{\alpha _t}}}}{{{P_t}{C_{BR}}C_{RU}^{rfr}{\beta _{rfr}}\beta }}} \right)^{\alpha  + n}}\frac{{{R^{{\alpha _t}\left( {\alpha  + n} \right) + 2}}}}{{ {{\alpha _t}\left( {\alpha  + n} \right) + 2} }}.
\end{align}
\begin{IEEEproof}
The proof is similar to the Appendix~D.
\end{IEEEproof}
\end{theorem}

\vspace{-0.7cm}
\section{Asymptotic Outage Performance and Diversity Analysis }
\vspace{-0.2cm}

Based on our analysis, we note that the two models, namely the central limit model and the curve fitting model, match the outage probability perfectly when the transmit power is not too large, while the performance in high SNR regions is not matched well. When we verify the aforementioned two models, we find that the two channel models perform as upper or lower limits when the transmit SNR is high enough. This is because both the central limit theorem and the curve fitting tool have changed the distribution of the STAR-IOS channels. Hence, we cannot obtain the accurate diversity order but receive the upper or lower boundaries of the accurate diversity orders. To obtain the accurate ones, we first exploit the M-fold convolution model to calculate the asymptotic expressions of the outage probability for the NOMA users under three protocols, i.e., the ES, MS, and TS protocols. Then, we harness the asymptotic expressions to derive and compare the accurate diversity orders among the three protocols.

In the following, we derive the asymptotic expressions by the M-fold convolution model based on the ES, MS and TS protocols, respectively, which are shown as \textbf{Theorem \ref{mfcml}} to \textbf{Theorem \ref{mfcmr}} and \textbf{Corollary \ref{MSl}} to \textbf{Corollary \ref{TSr}}. We then compare the diversity orders of three protocols via the M-fold convolution model as \textbf{Corollary \ref{diversityorder1}} to \textbf{Corollary \ref{diversityorder4}} and \textbf{Remark \ref{r3}} to \textbf{Remark \ref{diversityorderend}}.
\vspace{-0.5cm}
\subsection{Asymptotic Analysis on the M-Fold Convolution Model}
\vspace{-0.2cm}
Based on the ES protocol, we first derive the closed-form asymptotic outage probability expressions for the reflecting user and the transmitting user as \textbf{Theorem \ref{mfcml}} and \textbf{Theorem \ref{mfcmr}}, respectively. Then, we derive the asymptotic outage expressions based on the MS and TS protocols by \textbf{Corollary \ref{MSl}} and \textbf{Corollary \ref{TSr}} in the following.

\subsubsection{The Reflecting User}
 Considering the ES, MS, and TS protocols, we first derive the asymptotic expressions of outage probability for the reflecting users in the following.

\begin{theorem}\label{mfcml}
We note that all the channels for different STAR-IOS elements are independent. Additionally, it is tractable to calculate that the outage situation always happens for the case as ${a_{rfr}} < \gamma _{th}^{SIC}{a_{rfl}}$. Under the ES protocol and considering ${a_{rfr}} > \gamma _{th}^{SIC}{a_{rfl}}$, we derive the closed-form asymptotic outage probability expression for the transmitting user as
\begin{align}
P_{out,rfl}^{mf,ES}\left( x \right) = \frac{{{{\left[ {\sigma \left( {0,0} \right)} \right]}^M}{R^{{\alpha _t}M}}}}{{M\left( {{\alpha _t}M + 2} \right)\left( {2M - 1} \right)!{{\left( {{\beta _{rfl}}} \right)}^M}}}{\left( {\frac{{{\Upsilon _{\max }}d_{BR}^{{\alpha _t}}}}{{{P_t}{C_{BR}}C_{RU}^{rfl}}}} \right)^M}.
\end{align}
\begin{IEEEproof}
As we have derived the CDF of the STAR-IOS channel model via the M-fold convolution method in \textbf{Lemma \ref{channel2}}, we substitute the CDF of the M-fold convolution model into the equation \eqref{Poutdef_rfl}, we obtain the integration as
\begin{align}
P_{out,rfl}^{mf,ES} (x)=& \int_0^R {{F_{{{\left| {g_m^{rfl}} \right|}^2}}}\left( {\frac{{{\Upsilon _{\max }}d_{BR}^{{\alpha _t}}{x^{{\alpha _t}}}}}{{{P_t}{C_{BR}}C_{RU}^{rfl}}}} \right)} {f_{d_{RU}^{rfl}}}\left( x \right)dx\notag\\
=& \frac{{{{\left[ {\sigma \left( {0,0} \right)} \right]}^M}}}{{M\left( {2M - 1} \right)!{{\left( {{\beta _{rfl}}} \right)}^M}}}{\left( {\frac{{{\Upsilon _{\max }}d_{BR}^{{\alpha _t}}}}{{{P_t}{C_{BR}}C_{RU}^{rfl}}}} \right)^M}\int_0^R {\frac{{{x^{{\alpha _t}M + 1}}}}{{{R^2}}}} dx.
\end{align}

By calculating the integration $\int_0^R {\frac{{{x^{{\alpha _t}M + 1}}}}{{{R^2}}}} dx = \frac{{{R^{{\alpha _t}M}}}}{{{\alpha _t}M + 2}}$, we derive the final expressions.
\end{IEEEproof}
\end{theorem}

\begin{corollary}\label{MSl}
For the MS protocol, the asymptotic outage probability expression is derived as
\begin{align}
P_{out,rfl}^{mf,MS}\left( x \right) = \frac{{{{\left[ {\sigma \left( {0,0} \right)} \right]}^{{M_{rfl}}}}{R^{{\alpha _t}{M_{rfl}}}}}}{{{M_{rfl}}\left( {{\alpha _t}{M_{rfl}} + 2} \right)\left( {2{M_{rfl}} - 1} \right)!}}{\left( {\frac{{{\Upsilon _{\max }}d_{BR}^{{\alpha _t}}}}{{{P_t}{C_{BR}}C_{RU}^{rfl}}}} \right)^{{M_{rfl}}}}.
\end{align}
\begin{IEEEproof}
The proof is similar to \textbf{Theorem \ref{mfcml}}.
\end{IEEEproof}
\end{corollary}

\begin{corollary}\label{TSl}
For the TS protocol, we derive the closed-form asymptotic outage probability expression as
\begin{align}
P_{out,rfl}^{mf,TS}\left( x \right) = \frac{{{{\left[ {\sigma \left( {0,0} \right)} \right]}^M}{R^{{\alpha _t}M}}}}{{M\left( {{\alpha _t}M + 2} \right)\left( {2M - 1} \right)!}}{\left( {\frac{{{\Upsilon _{\max }}d_{BR}^{{\alpha _t}}}}{{{P_t}{C_{BR}}C_{RU}^{rfl}}}} \right)^M}.
\end{align}
\begin{IEEEproof}
The proof is similar to \textbf{Theorem \ref{mfcml}}.
\end{IEEEproof}
\end{corollary}

\subsubsection{The Transmitting User}
Then, we derive the asymptotic outage probability expressions of the transmitting user based on the ES, MS, and TS protocols in the following.

\begin{theorem}\label{mfcmr}
Note that the outage probability for the transmitting user is constantly equal to one when ${a_{rfr}} < \gamma _{th}{a_{rfl}}$. Hence, we consider ${a_{rfr}} > \gamma _{th}{a_{rfl}}$ and derive the closed-form asymptotic outage probability expression for the transmitting user as
\begin{align}
P_{out,rfr}^{mf,ES}\left( x \right) = \frac{{{{\left[ {\sigma \left( {0,0} \right)} \right]}^M}{R^{{\alpha _t}M}}}}{{M\left( {{\alpha _t}M + 2} \right)\left( {2M - 1} \right)!{{\left( {{\beta _{rfr}}} \right)}^M}}}{\left( {\frac{{{\Upsilon _2}d_{BR}^{{\alpha _t}}}}{{{P_t}{C_{BR}}C_{RU}^{rfr}}}} \right)^M}.
\end{align}
\begin{IEEEproof}
We substitute the CDF of the M-fold convolution model in \textbf{Lemma \ref{channel2}} into \eqref{Poutdef_rfr}, we express the outage probability of the transmitting user as
\begin{align}
P_{out,rfr}^{mf,ES} = \frac{{{{\left[ {\sigma \left( {0,0} \right)} \right]}^M}}}{{{{\left( {{\beta _{rfr}}} \right)}^M}M\left( {2M - 1} \right)!}}{\left( {\frac{{{\Upsilon _{\rm{2}}}d_{BR}^{{\alpha _t}}}}{{{P_t}{C_{BR}}C_{RU}^{rfr}}}} \right)^M}\int_0^R {\frac{{{x^{{\alpha _t}M + 1}}}}{{{R^2}}}dx} ,
\end{align}
and this integration can be easily calculated to obtain the final expression.
\end{IEEEproof}
\end{theorem}

\begin{corollary}\label{MSr}
For the MS protocol, the asymptotic outage probability expressions for the transmitting user is derived as
\begin{align}
P_{out,rfr}^{mf,MS}\left( x \right) = \frac{{{{\left[ {\sigma \left( {0,0} \right)} \right]}^{{M_{rfr}}}}{R^{{\alpha _t}{M_{rfr}}}}}}{{{M_{rfr}}\left( {{\alpha _t}{M_{rfr}} + 2} \right)\left( {2{M_{rfr}} - 1} \right)!}}{\left( {\frac{{{\Upsilon _2}d_{BR}^{{\alpha _t}}}}{{{P_t}{C_{BR}}C_{RU}^{rfr}}}} \right)^{{M_{rfr}}}}.
\end{align}
\begin{IEEEproof}
The proof is similar to \textbf{Theorem \ref{mfcmr}}.
\end{IEEEproof}
\end{corollary}

\begin{corollary}\label{TSr}
For the TS protocol, the asymptotic outage probability expressions for the transmitting user is derived as
\begin{align}
P_{out,rfr}^{mf,TS}\left( x \right) = \frac{{{{\left[ {\sigma \left( {0,0} \right)} \right]}^M}{R^{{\alpha _t}M}}}}{{M\left( {{\alpha _t}M + 2} \right)\left( {2M - 1} \right)!}}{\left( {\frac{{{\Upsilon _2}d_{BR}^{{\alpha _t}}}}{{{P_t}{C_{BR}}C_{RU}^{rfr}}}} \right)^M}.
\end{align}
\begin{IEEEproof}
As the proof has provided in \textbf{Theorem \ref{mfcmr}}, we omit the proof.
\end{IEEEproof}
\end{corollary}

\subsection{Diversity Analysis}
After deriving the asymptotic expressions for the ES, MS, and TS protocols, we then derive the diversity orders  for the reflecting user and the transmitting user as the following remarks and corollaries.

\subsubsection{The ES Protocol}
We firstly express the diversity orders of the reflecting and transmitting users based on the ES protocol as follows.

\begin{corollary}\label{diversityorder1}
We assume the transmit SNR $\rho_t \to \infty  $. For the ES protocol, the accurate diversity order of the reflecting user is derived as
\begin{align}
{d_{rfl}^{ES}} =  - \mathop {\lim }\limits_{{\rho _t} \to \infty } \frac{{\log \left[ {P_{out,rfl}^{mf,ES}\left( {{\rho _t}} \right)} \right]}}{{\log \left( {{\rho _t}} \right)}}  = M.
\end{align}
where $\rho_t = P_t/\sigma^2$
\begin{IEEEproof}
As ${P_{out,rfl}^{mf,ES}\left( {{\rho _t}} \right)}$ is expressed as ${P_{out,rfl}^{mf,ES}\left( {{\rho _t}} \right)} = A\rho_t^{-M}$, where $A$ is the constant without $\rho_t$. Hence, we calculate the limit as $ {d_{rfl}} =  - \mathop {\lim }\limits_{{\rho _t} \to \infty } \frac{{\log \left[ {A\rho _t^{ - M}} \right]}}{{\log \left( {{\rho _t}} \right)}} = M$ and the proof is end.
\end{IEEEproof}
\end{corollary}

\begin{corollary}\label{diversityorder2}
We assume the transmit SNR $\rho_t \to \infty  $. For the ES protocol, the accurate diversity order of the reflecting user be derived as
\begin{align}
{d_{rfr}^{ES}} =  - \mathop {\lim }\limits_{{\rho _t} \to \infty } \frac{{\log \left[ {P_{out,rfr}^{mf,ES}\left( {{\rho _t}} \right)} \right]}}{{\log \left( {{\rho _t}} \right)}} = M.
\end{align}
\begin{IEEEproof}
The proof is similar to \textbf{Corollary \ref{diversityorder1}}.
\end{IEEEproof}
\end{corollary}

\begin{remark}\label{r3}
For the ES protocol, the diversity orders for the reflecting and transmitting user are equal to $M$, which is the total number of the STAR-IOS elements.
\end{remark}

\subsubsection{The MS Protocol}
We then derive the diversity orders of the reflecting and transmitting users based on the MS protocol by the following corollary and remark.

\begin{corollary}\label{diversityorder3}
We assume the transmit SNR $\rho_t \to \infty  $. For the MS protocol, we derive the accurate diversity orders of the NOMA users as
\begin{align}
d_{rfl}^{MS} =  - \mathop {\lim }\limits_{{\rho _t} \to \infty } \frac{{\log \left[ {P_{out,rfl}^{mf,MS}\left( {{\rho _t}} \right)} \right]}}{{\log \left( {{\rho _t}} \right)}} = {M_{rfl}},\\
d_{rfr}^{MS} =  - \mathop {\lim }\limits_{{\rho _t} \to \infty } \frac{{\log \left[ {P_{out,rfr}^{mf,MS}\left( {{\rho _t}} \right)} \right]}}{{\log \left( {{\rho _t}} \right)}} = {M_{rfr}}.
\end{align}
\begin{IEEEproof}
The proof is similar to \textbf{Corollary \ref{diversityorder1}}.
\end{IEEEproof}
\end{corollary}

\begin{remark}
For the MS protocol, the diversity orders for the reflecting and transmitting user are equal to $M_{rf}$, where $rf \in \{rfl,rfr\}$ for the reflecting links and transmitting links, respectively. This value is the active number of the STAR-IOS elements.
\end{remark}

\subsubsection{The TS Protocol}
We additionally derive the diversity orders of the reflecting and transmitting users according to the TS protocol.

\begin{corollary}\label{diversityorder4}
We assume the transmit SNR $\rho_t \to \infty  $. For the TS protocol, we calculate the diversity orders for the users in the NOMA pair as
\begin{align}
d_{rfl}^{TS} = d_{rfr}^{TS} = M.
\end{align}
which is the number of the total STAR-IOS elements.
\end{corollary}

\begin{remark}\label{diversityorderend}
When we compare the diversity orders of the three protocols in Table \ref{table1}, we conclude that the diversity orders of users are equal to the number of active STAR-IOS elements. For the ES and TS protocols, all the STAR-IOS elements are fully activated, thereby the diversity orders are large. However, for the MS protocol, the active STAR-IOS elements are split into two portions to be exploited to reflect or refract signals. Hence, the diversity orders of the MS protocol generally cannot match the transmitting and reflecting gain of the ES or TS protocol.
\begin{table}
 \vspace{-0.2cm}
 \caption{Diversity orders for different STAR-IOS protocols.}
 \vspace{-0.2cm}
\label{table1}
\centering
\begin{tabular}{|c|c|c|c|}
\hline
  Protocol & The TS Protocol& The ES Protocol& The MS Protocol\\
    \hline
  Diversity Orders & $M$ & $M$& $M_{rf}$ \\
  \hline
\end{tabular}
\end{table}
\vspace{-0.4cm}
\end{remark}

\section{Numerical Results}

This section presents the numerical results for the outage performance of users. More specifically, the Monte Carlo simulations validate 1) the analytical closed-form expressions based on the central limit model and the curve fitting model, and 2) the diversity orders based on the M-fold convolution model with three protocols, namely the ES, MS, and TS protocols. Unless otherwise stated, the numerical coefficients are defined as follows. The noise power is ${\sigma ^2} =  - 170 + 10\log \left( {f_c} \right) + {N_f}=-90 $ dB, where $N_f$ is $10$ dB and the carrier bandwidth $f_c$ is $10$ MHz. The transmit power $P_t$ varies in $\left[10,24\right]$ dBm. The path loss exponent is $\alpha_t = 2.4$. The outage threshold and the SIC threshold are equal as $\gamma_{th}^{SIC}=\gamma_{th}^{out}=2^{0.1}-1$. The power allocation coefficients of the BS are $a_{rfr} = 0.6$ and $ a_{rfl} = 0.4$. The energy splitting coefficients for the ES protocol are $\beta_{rfr}=0.3$ and $\beta_{rfl}=0.7$. The distance from the BS to STAR-IOS is $d_{BR}=100$ m. The other coefficients are varied and specifically defined in the following paragraphs. Additionally, we denote ``CL" as the central limit model and ``CF" as the curve fitting model in the legends.

\begin{figure}[htbp]
\centering    
\vspace{-0.5cm}
\subfigure[] 
{
	\begin{minipage}{7cm}
	\centering          
	\includegraphics[width= 3in]{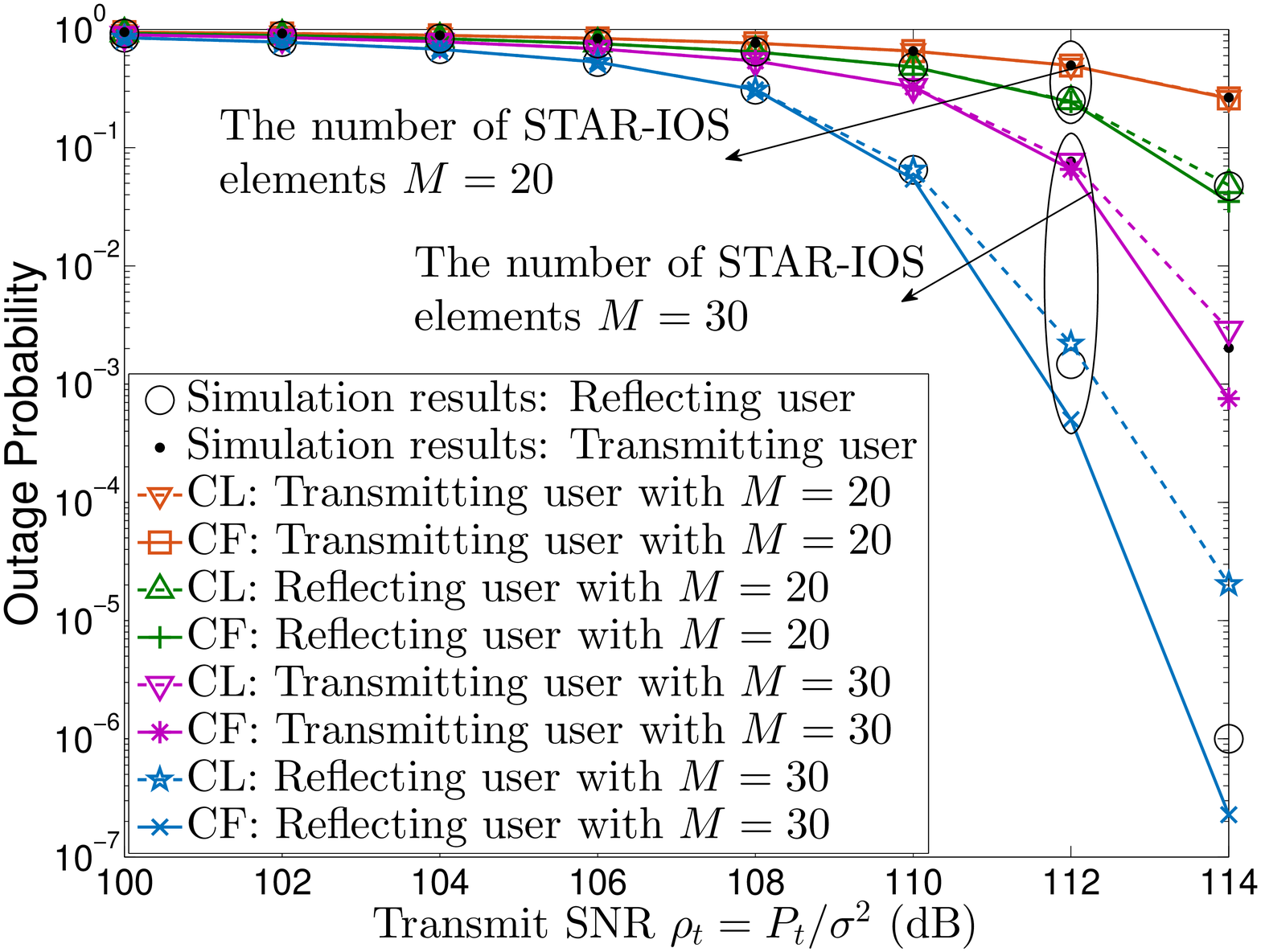}   
\vspace{-0.3cm}
    \label{figure3}
	\end{minipage}
}
\subfigure[] 
{
	\begin{minipage}{7cm}
	\centering      
	\includegraphics[width= 3in]{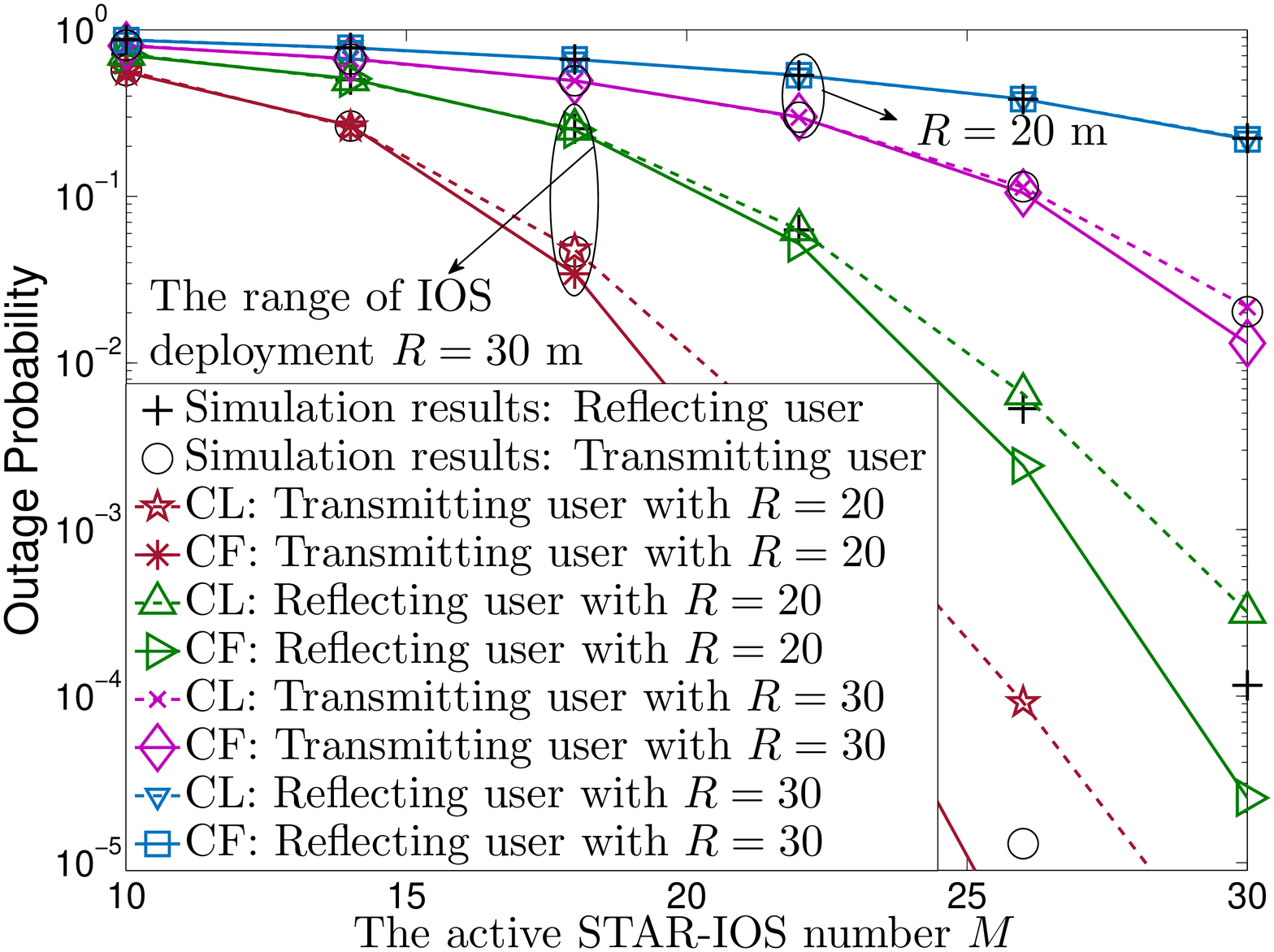}   
\vspace{-0.3cm}
    \label{figure4}
	\end{minipage}
}
\vspace{-0.3cm}
\caption{Comparisons between the curve fitting model and the central limit model based on the ES protocol: (a) Outage probability versus the transmit SNR with various numbers of STAR-IOS elements $M=\{20,30\}$; (b) Outage probability versus the number of STAR-IOS elements with various STAR-IOS deployment radii $R=\{20,30\}$ m.} 
\vspace{-0.4cm}
\end{figure}

In Fig. \ref{figure3}, we further define the number of STAR-IOS elements as $M=\{20,30\}$ and the radius of STAR-IOS deployment area as $R=20$ m. Then, we investigate the outage performance versus the transmit SNR $\rho_t = P_t/\sigma^2$. One observation is that increasing the number of STAR-IOS elements is able to significantly increase the performance of NOMA users, which is because a large number of STAR-IOS elements provide well-integrated signals to enhance the channel quality.

In Fig. \ref{figure4}, the coefficients are defined that the number of STAR-IOS elements varies in $M=[10,30]$ and the radius of STAR-IOS deployment area is chosen from $R=\{20,30\}$ m. Hence, the outage performance versus the number of STAR-IOS elements is evaluated. We observe that reducing the STAR-IOS deployment range enhances the performance as it reduces the influence of path loss. Comparing Fig. \ref{figure3} and Fig. \ref{figure4}, we conclude that the two channel models match the low SNR regions better than the high SNR regions. Specifically, in high SNR regions, the curve fitting model performs as a lower bound of the simulation results while the central limit model is an upper bound.

\begin{figure}[htbp]
\centering    
\vspace{-0.3cm}
\subfigure[] 
{
	\begin{minipage}{7cm}
	\centering          
	\includegraphics[width= 3in]{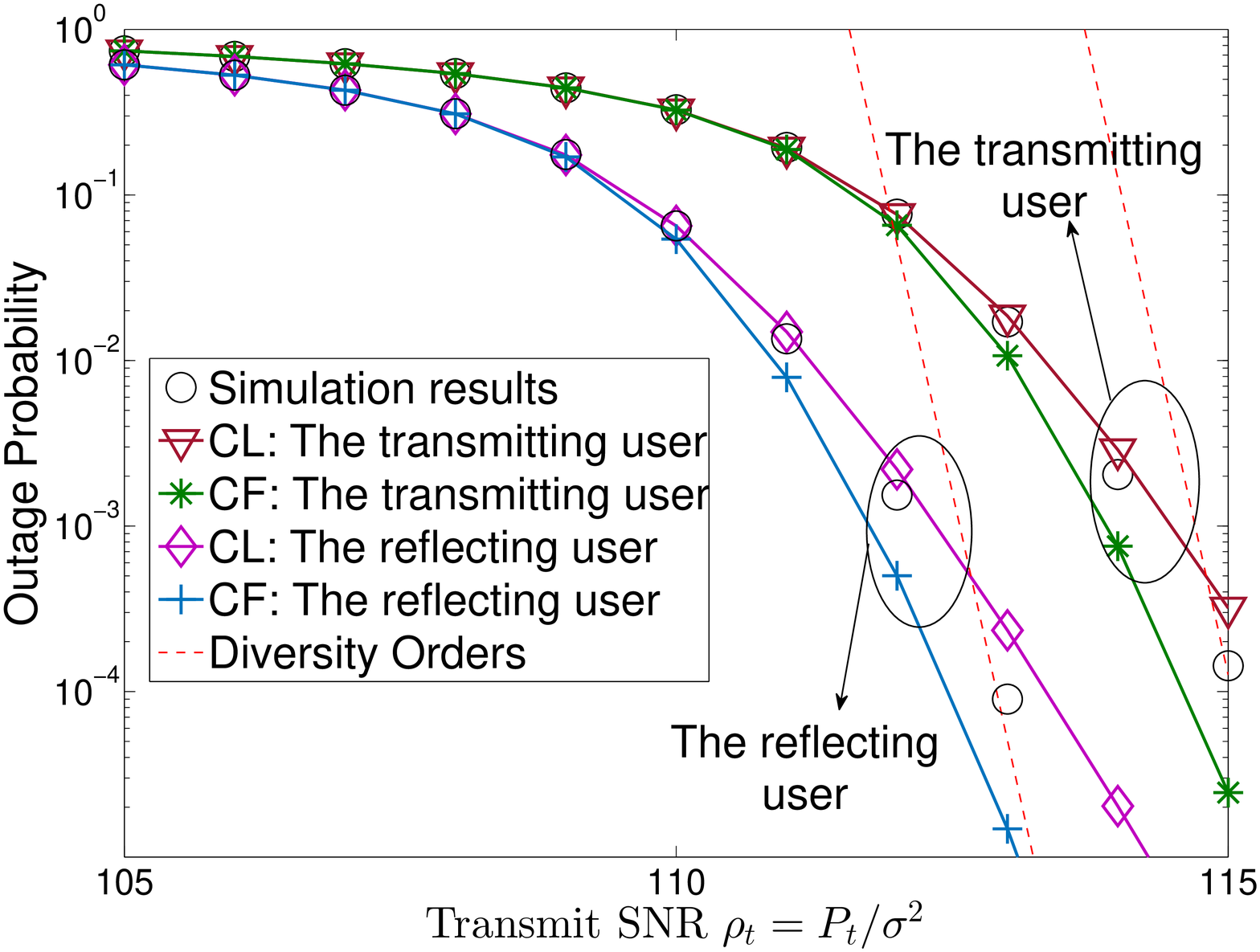}   
\vspace{-0.3cm}
    \label{figure5}
	\end{minipage}
}
\subfigure[] 
{
	\begin{minipage}{7cm}
	\centering      
	\includegraphics[width= 3in]{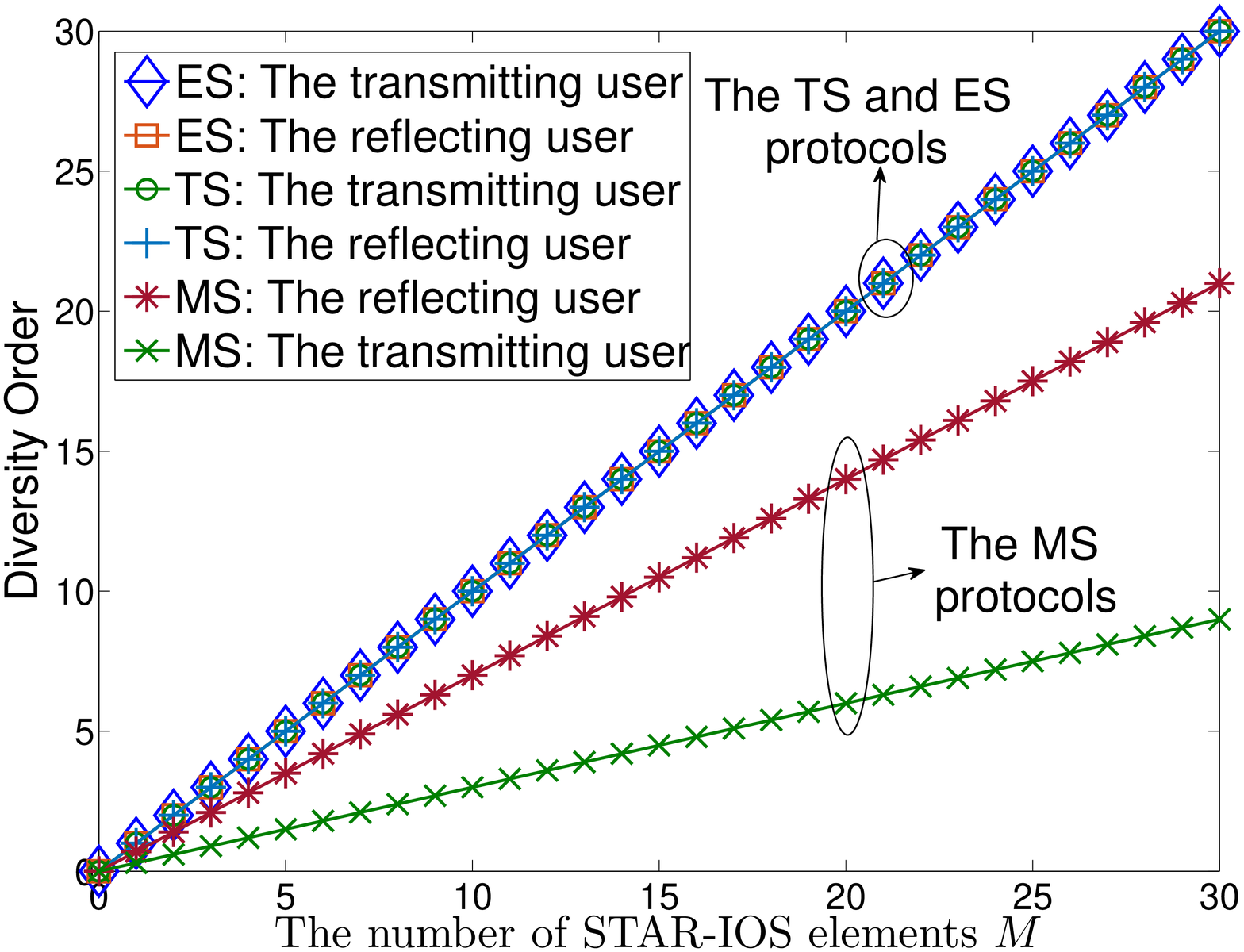}   
\vspace{-0.3cm}
    \label{figure6}
	\end{minipage}
}
\vspace{-0.3cm}
\caption{Validating and simulations for diversity orders: (a) Outage probability versus the transmit SNR for different channel models, i.e., the central limit model, the curve fitting model, and the M-fold convolution model; (b) Diversity gains versus the number of STAR-IOS elements to compare different protocols, i.e., the ES protocol, the TS protocol, and the MS protocol.} 
\vspace{-0.6cm}
\end{figure}

With $M = 30$ and $R=20$ m, the Fig. \ref{figure5} validates the correction of diversity orders derived by the M-fold convolution model. When we consider that in the MS protocol, $70\rm{\% }$ of STAR-IOS elements are utilized for the reflecting user and the rest are exploited for the transmitting user, the Fig. \ref{figure6} indicates the diversity gains versus the active number of STAR-IOS elements. Based on the two figures, we observe that the diversity orders are the same as the active STAR-IOS elements, while the diversity order for the MS protocol is lower than that for the TS and ES protocols, which fits our analytical diversity gains. This is because the MS protocol have fewer active STAR-IOS elements than other protocols.

\begin{figure}[htbp]
\centering    
\vspace{-0.4cm}
\subfigure[] 
{
	\begin{minipage}{7cm}
	\centering          
	\includegraphics[width= 3in]{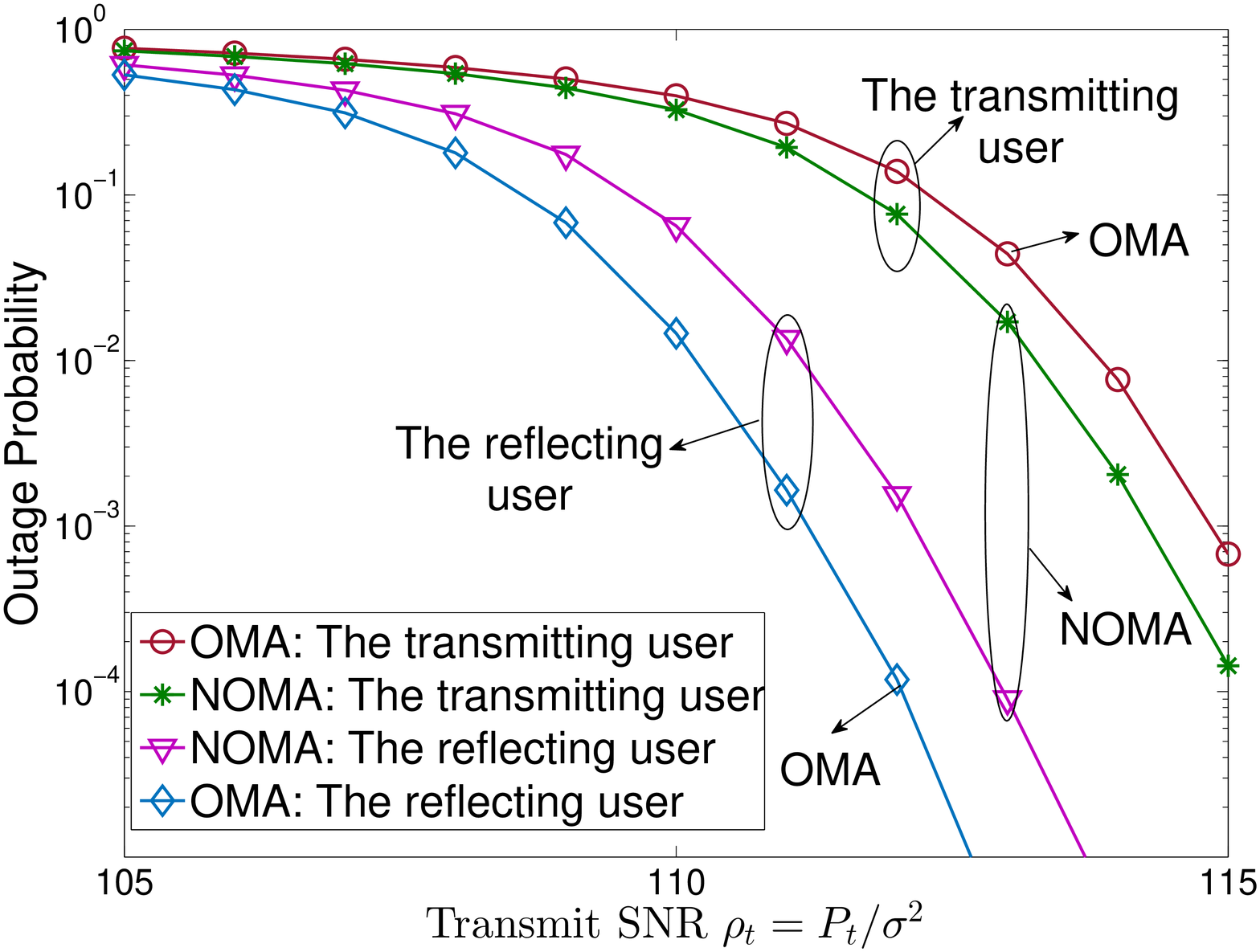}   
\vspace{-0.3cm}
    \label{figure7}
	\end{minipage}
}
\subfigure[] 
{
	\begin{minipage}{7cm}
	\centering      
	\includegraphics[width= 3in]{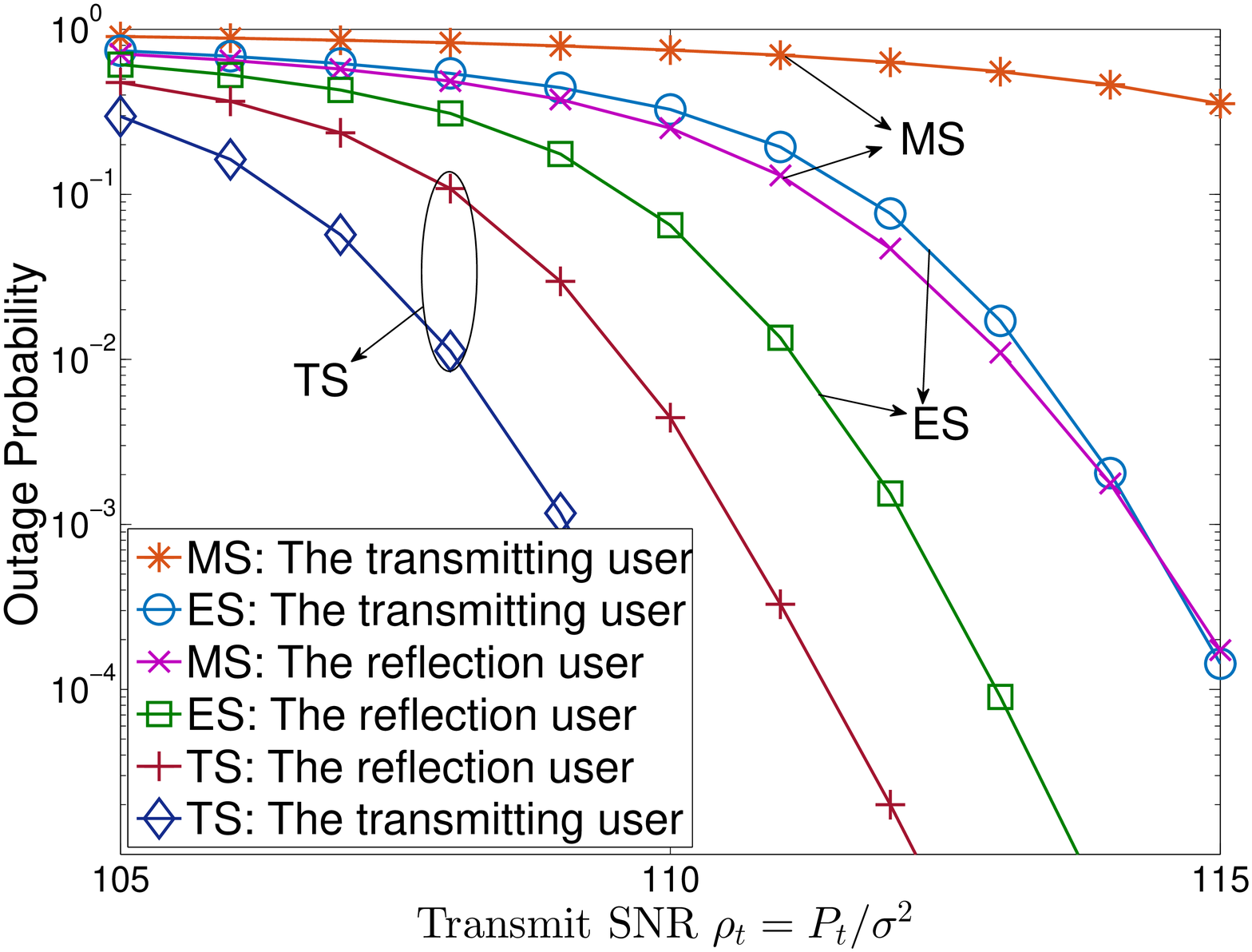}   
\vspace{-0.3cm}
    \label{figure8}
	\end{minipage}
}
\vspace{-0.3cm}
\caption{Simulations for outage probability: (a) Outage probability versus the transmit SNR to compare the NOMA technique and the OMA technique; (b) Outage probability versus the transmit SNR to compare different protocols, i.e., the ES protocol, the TS protocol, and the MS protocol.} 
\vspace{-0.5cm}
\end{figure}

With the same setting of Fig. \ref{figure5} and \ref{figure6}, we compare the outage performance of the NOMA and OMA techniques in Fig. \ref{figure7}. We additionally compare the outage performance of the ES, TS, and MS protocols versus the transmit SNR in Fig. \ref{figure8}. One observation is that the TS protocol performs best but it cannot serve two users in the same time block. To serve two users in the same resource block, the ES protocol outperforms the MS protocol as the ES protocol integrates more STAR-IOS elements for each user to achieve higher diversity gains than the MS protocol.

\vspace{-0.4cm}

\section{Conclusions}

This paper has proposed three channel models of STAR-IOS channels and evaluated the outage performance of a STAR-IOS-aided downlink NOMA framework with randomly deployed users. More specifically, we have exploited the central limit model and the curve fitting model to derive the closed-form outage probability expressions and exploited the M-fold convolution model to derive the diversity gains under the ES, MS, and TS protocols, respectively. The analytical results have revealed that 1) the central limit model has the closed-form expression as the manipulation of error functions; 2) the coefficients of the curve fitting model have the relationship with the number of STAR-IOS elements as $\alpha = M$ and $\beta<M$; and 3) the diversity gains under three protocols are equal to the active number of STAR-IOS elements. Numerical results have shown that: 1) STAR-IOSs enhance the channel quality of its aided user; 2) the central limit model and the curve fitting model perform as boundaries of the simulation results in high SNR regions; 3) the TS protocol has the best performance but it cannot serve two users in the same time block; and 4) with two users in the same resource block, the ES protocol outperforms the MS protocols.
\vspace{-0.3cm}
\section*{Appendix~A: Proof of Lemma~\ref{channel1}} \label{Appendix:A}
\renewcommand{\theequation}{A.\arabic{equation}}
\setcounter{equation}{0}

As ${\left| {g_m^{rf}} \right|^2} = {\beta _{rf}}{\left( {\sum\limits_{m = 1}^M {h_{RU,m}^{rf}{h_{BR,m}}} } \right)^2}$ for $rf \in \left\{ {rfl,rfr} \right\}$, we exploit the central limit theorem to derive the variable $\left| {g_m^{rf}} \right|=\sqrt {{\beta _{rf}}} \sum\limits_{m = 1}^M {h_{RU,m}^{rf}{h_{BR,m}}} $. As we assume that independent Rician variables have the same mean and variance, denoted as $\bar h$ and $\eta$, we derive the mean and variance of $\left| {g_m^{rf}} \right|$ as
\begin{align}\label{Eeq}
\bar h_{eq}^{rf} &= E\left( {\left| {g_m^{rf}} \right|} \right) = E\left( {\sqrt {{\beta _{rf}}} \sum\limits_{m = 1}^M {h_{RU,m}^{rf}{h_{BR,m}}} } \right) = \sqrt {{\beta _{rf}}} M{{\bar h}^2},\\
\label{Veq}
\eta _{eq}^{rf} &= Var\left( {\left| {g_m^{rf}} \right|} \right) = Var\left( {\sqrt {{\beta _{rf}}} \sum\limits_{m = 1}^M {h_{RU,m}^{rf}{h_{BR,m}}} } \right) = {\beta _{rf}}M\left( {2{{\bar h}^2}\eta  + {\eta ^2}} \right),
\end{align}
where $E(\cdot)$ and $Var(\cdot)$ are the expectation and variance of a certain variable.

Hence, we express the PDF of $\left| {g_m^{rf}} \right|$ as
\begin{align}
{f_{\left| {g_m^{rf}} \right|}}\left( x \right) = \frac{1}{{\sqrt {2\pi \eta _{eq}^{rf}} }}\exp \left( { - \frac{{{{\left( {x - \bar h_{eq}^{rf}} \right)}^2}}}{{2\eta _{eq}^{rf}}}} \right).
\end{align}

Based on the derivation for a variable $x$ that ${f_{{x^2}}}\left( y \right) = \frac{1}{{2\sqrt y }}\left( {{f_x}\left( {\sqrt y } \right) + {f_x}\left( { - \sqrt y } \right)} \right)$, we derive the PDF of ${{{\left| {g_m^{rf}} \right|}^2}}$ as
\begin{align}
f_{{{\left| {g_m^{rf}} \right|}^2}}\left( y \right) = \frac{1}{{{\rm{2}}\sqrt {2\pi \eta _{eq}^{rf}y} }}\left( {\exp \left( { - \frac{{{{\left( {\sqrt y  - \bar h_{eq}^{rf}} \right)}^2}}}{{2\eta _{eq}^{rf}}}} \right){\rm{ + }}\exp \left( { - \frac{{{{\left( {\sqrt y {\rm{ + }}\bar h_{eq}^{rf}} \right)}^2}}}{{2\eta _{eq}^{rf}}}} \right)} \right).
\end{align}

Then, we derive the CDF of the channel model as \textbf{Lemma \ref{channel1}} according to the derivation as
$ \int_0^x {\frac{1}{{\sqrt y }}\left( {\exp \left( { - \frac{{{{\left( {\sqrt y  - a} \right)}^2}}}{b}} \right){\rm{ + }}\exp \left( { - \frac{{{{\left( {\sqrt y {\rm{ + }}a} \right)}^2}}}{b}} \right)} \right)dy = } \sqrt {\pi b} \left( {\erf\left( {\frac{{a + \sqrt x }}{{\sqrt b }}} \right) - \erf\left( {\frac{{a - \sqrt x }}{{\sqrt b }}} \right)} \right)$.
\vspace{-0.3cm}
\section*{Appendix~B: Proof of Lemma~\ref{channel2}} \label{Appendix:B}
\renewcommand{\theequation}{B.\arabic{equation}}
\setcounter{equation}{0}

With the assumption that $k_1$ and $k_2$ represent the Rician coefficients for the BR links and RU links, respectively, we obtain that the PDF expression of $|{h_{RU,m}^{rf}{h_{BR,m}}}|$ is expressed as ${f_{|h_{RU,m}^{rf}{h_{BR,m}}|}}\left( z \right) = \int_0^\infty  {\frac{1}{w}} {f_{|h_{RU,m}^{rf}|}}\left( w \right){f_{|{h_{BR,m}}|}}\left( {\frac{z}{w}} \right)dw$. Exploiting the series of the Bessel function as ${I_v}\left[ p \right] = \sum\limits_{s = 0}^\infty  {\frac{1}{{s!\Gamma \left( {s + v + 1} \right)}}} {\left( {\frac{p}{2}} \right)^{2s + v}}$, we derive ${f_{|h_{RU,m}^{rf}{h_{BR,m}}|}}$ as
\begin{align}
{f_{\left| {h_{RU,m}^{rf}{h_{BR,m}}} \right|}}\left( z \right) =& \frac{{4\left( {1 + {k_{\rm{1}}}} \right)\left( {1 + {k_{\rm{2}}}} \right){z^{2n + 1}}}}{{\exp \left( {{k_{\rm{1}}}} \right)\exp \left( {{k_{\rm{2}}}} \right)}}\sum\limits_{t = 0}^\infty  {\frac{{{{\left( {{k_{\rm{1}}}\left( {1 + {k_{\rm{1}}}} \right)} \right)}^t}}}{{t!\Gamma \left( {t + 1} \right)}}\sum\limits_{n = 0}^\infty  {\frac{{{{\left( {{k_{\rm{2}}}\left( {1 + {k_{\rm{2}}}} \right)} \right)}^n}}}{{n!\Gamma \left( {n + 1} \right)}}} } \notag\\
&\times\int_0^\infty  {\frac{{{x^{t - n - 1}}}}{2}} \exp \left[ { - \left( {1 + {k_{\rm{1}}}} \right)x} \right]\exp \left[ { - \left( {1 + {k_2}} \right)\frac{{{z^2}}}{x}} \right]dw.
\end{align}

Based on Eq. [2.3.16.1] in \cite{table} and formula calculation, the PDF is further derived as
\begin{align}\label{b2}
{f_{\left| {h_{RU,m}^{rf}{h_{BR,m}}} \right|}}\left( z \right) = & 4{z^{t + n + 1}}\sum\limits_{t = 0}^\infty  {\sum\limits_{n = 0}^\infty  {\frac{{k_1^tk_2^t{{\left[ {\left( {1 + {k_{\rm{1}}}} \right)\left( {1 + {k_{\rm{2}}}} \right)} \right]}^{\frac{{t + n}}{2} + 1}}}}{{{{\left( {t!} \right)}^2}{{\left( {n!} \right)}^2}\exp \left( {{k_{\rm{1}}} + {k_{\rm{2}}}} \right)}}} } \notag\\
&\times {K_{t - n}}\left( {2z\sqrt {\left( {1 + {k_{\rm{1}}}} \right)\left( {1 + {k_{\rm{2}}}} \right)} } \right),
\end{align}
where ${K_v}\left(  \cdot  \right)$ is the modified Bessel function of the second kind.

Based on \eqref{b2}, we express the Laplace transform expression of ${\left| {h_{RU,m}^{rf}{h_{BR,m}}} \right|}$ as
\begin{align}
{\cal L}\left[ {{f_{\left| {h_{RU,m}^{rf}{h_{BR,m}}} \right|}}\left( x \right)} \right] =& 4\sum\limits_{t = 0}^\infty  {\sum\limits_{n = 0}^\infty  {\frac{{k_1^tk_2^t{{\left[ {\left( {1 + {k_{\rm{1}}}} \right)\left( {1 + {k_{\rm{2}}}} \right)} \right]}^{\frac{{t + n}}{2} + 1}}}}{{{{\left( {t!} \right)}^2}{{\left( {n!} \right)}^2}\exp \left( {{k_{\rm{1}}} + {k_{\rm{2}}}} \right)}}} }\notag \\
&\times \int_0^\infty  {{x^{t + n + 1}}\exp \left( { - sx} \right)} {K_{t - n}}\left( {2x\sqrt {\left( {1 + {k_{\rm{1}}}} \right)\left( {1 + {k_{\rm{2}}}} \right)} } \right).
\end{align}

Based on M-fold convolution, we express the Laplace transform expression of the sum of combined channels of different STAR-IOS elements $\sum\limits_{m = 1}^M |{h_{RU,m}^{rf}{h_{BR,m}}}|$ as
\begin{align}\label{b3}
\mathcal{L}\left[ {\left| {\sum\limits_{m = 1}^M {h_{RU,m}^{rf}{h_{BR,m}}} } \right|} \right]\left( s \right) = {\left\{ {\mathcal{L}\left[ {{f_{\left| {h_{RU,m}^{rf}{h_{BR,m}}} \right|}}\left( x \right)} \right]} \right\}^M}.
\end{align}

Using Eq. [2.16.6.3] in \cite{table}, the Laplace transform of ${{f_{\left| {h_{RU,m}^{rf}{h_{BR,m}}} \right|}}\left( x \right)}$ is derived as
\begin{align}
&{\cal L}\left[ {{f_{\left| {h_{RU,m}^{rf}{h_{BR,m}}} \right|}}\left( x \right)} \right]\left( s \right) = 4\sum\limits_{t = 0}^\infty  {\sum\limits_{n = 0}^\infty  {\frac{{k_1^tk_2^t{{\left[ {\left( {1 + {k_{\rm{1}}}} \right)\left( {1 + {k_{\rm{2}}}} \right)} \right]}^{\frac{{t + n}}{2} + 1}}}}{{{{\left( {t!} \right)}^2}{{\left( {n!} \right)}^2}\exp \left( {{k_{\rm{1}}} + {k_{\rm{2}}}} \right)}}} }\notag \\
&\hspace*{3cm}\times \frac{{{{\left( {4\sqrt {\left( {1 + {k_1}} \right)\left( {1 + {k_2}} \right)} } \right)}^{t - n}}\sqrt \pi  }}{{{{\left( {s + 2\sqrt {\left( {1 + {k_1}} \right)\left( {1 + {k_2}} \right)} } \right)}^{2t + 2}}}}\Gamma \binom{2n + 2,2t + 2}{t + n + \frac{5}{2}}\notag\\
&\hspace*{3cm}\times {}_2{F_1}\left( {2t + 2,t - n + \frac{1}{2};t + n + \frac{5}{2};\frac{{s - 2\sqrt {\left( {1 + {k_1}} \right)\left( {1 + {k_2}} \right)} }}{{s + 2\sqrt {\left( {1 + {k_1}} \right)\left( {1 + {k_2}} \right)} }}} \right).
\end{align}

We assume $s \to \infty $. Then, the expression ${}_2{F_1}\left( {2t + 2,t - n + \frac{1}{2};t + n + \frac{5}{2};\frac{{s - 2\sqrt {\left( {1 + {k_1}} \right)\left( {1 + {k_2}} \right)} }}{{s + 2\sqrt {\left( {1 + {k_1}} \right)\left( {1 + {k_2}} \right)} }}} \right)$ is approximately expressed as ${}_2{F_1}\left( {2t + 2,t - n + \frac{1}{2};t + n + \frac{5}{2};1} \right)$. Hence, the Laplace transform is finally derived as
\begin{align}\label{b6}
{\cal L}\left[ {{f_{\left| {h_{RU,m}^{rf}{h_{BR,m}}} \right|}}\left( x \right)} \right]\left( s \right) =& \sum\limits_{t = 0}^\infty  {\sum\limits_{n = 0}^\infty  {\frac{{{4^{t - n + 1}}\sqrt \pi  k_1^tk_2^t{{\left[ {\left( {1 + {k_{\rm{1}}}} \right)\left( {1 + {k_{\rm{2}}}} \right)} \right]}^{t + 1}}}}{{{{\left( {t!} \right)}^2}{{\left( {n!} \right)}^2}\exp \left( {{k_{\rm{1}}} + {k_{\rm{2}}}} \right)}}} }\Gamma \binom{2n + 2,2t + 2}{t + n + \frac{5}{2}} \notag\\
&\times {}_2{F_1}\left( {2t + 2,t - n + \frac{1}{2};t + n + \frac{5}{2};1} \right){s^{ - (2t + 2)}}\notag\\
 =& \sum\limits_{t = 0}^\infty  {\sum\limits_{n = 0}^\infty  {\sigma \left( {t,n} \right)} } {s^{ - (2t + 2)}}.
\end{align}

Hence, substituting \eqref{b6} into \eqref{b3}, we derive the Laplace transform expression of $\sum\limits_{m = 1}^M |{h_{RU,m}^{rf}{h_{BR,m}}}|$ as
\begin{align}\label{b7}
{\cal L}\left[ {\left| {\sum\limits_{m = 1}^M {h_{RU,m}^{rf}{h_{BR,m}}} } \right|} \right]\left( s \right) = {\left\{ {{\cal L}\left[ {{f_{\left| {h_{RU,m}^{rf}{h_{BR,m}}} \right|}}\left( x \right)} \right]} \right\}^M} = {\left( {\sum\limits_{t = 0}^\infty  {\sum\limits_{n = 0}^\infty  {\sigma \left( {t,n} \right)} } {s^{ - (2t + 2)}}} \right)^M}.
\end{align}

We only keep the first item of the two Taylor series of the Bessel function in \eqref{b7}, which means we consider $n=0$ and $t=0$ in \eqref{b7}. We utilize the inverse Laplace transform to obtain the PDF of $\sum\limits_{m = 1}^M |{h_{RU,m}^{rf}{h_{BR,m}}}|$ as
\begin{align}
f_{\left| {\sum\limits_{m = 1}^M {h_{RU,m}^{rf}{h_{BR,m}}} } \right|}^{0 + }\left( x \right) = &{{\cal L}^{ - 1}}\left[ {{{\left( {\sum\limits_{t = 0}^\infty  {\sum\limits_{n = 0}^\infty  {\sigma \left( {t,n} \right)} } {s^{ - (2t + 2)}}} \right)}^M}} \right]\left( x \right)\notag\\
 =& {{\cal L}^{ - 1}}\left[ {{{\left( {\sigma \left( {0,0} \right){s^{ - 2}}} \right)}^M}} \right]\left( x \right)\notag\\
 =& \frac{{{{\left[ {\sigma \left( {0,0} \right)} \right]}^M}}}{{\left( {2M - 1} \right)!}}{x^{2M - 1}}.
\end{align}

Additionally, we note the equation ${f_{{X^2}}}\left( x \right) = \frac{1}{{2\sqrt x }}\left[ {{f_X}\left( {\sqrt x } \right) + {f_X}\left( { - \sqrt x } \right)} \right]$. We ignore the negative part of the aforementioned equation because of ${\left| {\sum\limits_{m = 1}^M {h_{RU,m}^{rf}{h_{BR,m}}} } \right|}>0$. And then, we derive the PDF of ${\left| {\sum\limits_{m = 1}^M {h_{RU,m}^{rf}{h_{BR,m}}} } \right|^2}$ as
\begin{align}
f_{{{\left| {\sum\limits_{m = 1}^M {h_{RU,m}^{rf}{h_{BR,m}}} } \right|}^2}}^{0 + }\left( x \right) = \frac{{{{\left[ {\sigma \left( {0,0} \right)} \right]}^M}{x^{M - 1}}}}{{2\left( {2M - 1} \right)!}}.
\end{align}

Note that we denote ${\left| {g_m^{rf}} \right|^2} = {\beta _{rf}}{\left| {\sum\limits_{m = 1}^M {h_{RU,m}^{rf}{h_{BR,m}}} } \right|^2},rf \in \left\{ {rfl,rfr} \right\}$. Hence, based on the equation ${f_{aX}}\left( x \right) = \frac{1}{{\left| a \right|}}{f_X}\left( {\frac{x}{a}} \right)$, we derive the PDF and CDF of ${\left| {g_m^{rf}} \right|^2}$ as the final expressions.

\vspace{-0.3cm}
\section*{Appendix~C: Proof of Theorem~\ref{clml}} \label{Appendix:C}
\renewcommand{\theequation}{C.\arabic{equation}}
\setcounter{equation}{0}

With the aid of the outage probability definition, the outage probability of the reflecting user is expressed as
\begin{align}\label{clm1}
P_{out,rfl}^{}\left( x \right) = \int_0^R {{F_{{{\left| {g_m^{rfl}} \right|}^2}}}\left( {\frac{{{\Upsilon _{\max }}d_{BR}^{{\alpha _t}}{x^{{\alpha _t}}}}}{{{P_t}{C_{BR}}C_{RU}^{rfl}}}} \right)} {f_{d_{RU}^{rfl}}}\left( x \right)dx.
\end{align}

Substituting the CDF of the central limit model and \eqref{rl} into \eqref{clm1}, we further derive the outage probability expression above as
\begin{align}\label{clm2}
P_{out,rfl}^{}\left( x \right)=\frac{1}{{{R^2}}}\int_0^R {x\left( {{\rm{erf}}\left( {\frac{{\bar h_{eq}^{rfl} + \sqrt {\frac{{{\Upsilon _{\max }}d_{BR}^{{\alpha _t}}{x^{{\alpha _t}}}}}{{{P_t}{C_{BR}}C_{RU}^{rfl}}}} }}{{\sqrt {2\eta _{eq}^{rf}} }}} \right) - {\rm{erf}}\left( {\frac{{\bar h_{eq}^{rfl} - \sqrt {\frac{{{\Upsilon _{\max }}d_{BR}^{{\alpha _t}}{x^{{\alpha _t}}}}}{{{P_t}{C_{BR}}C_{RU}^{rfl}}}} }}{{\sqrt {2\eta _{eq}^{rf}} }}} \right)} \right)}.
\end{align}

As the integration above cannot be derived, we utilize the Taylor series of the error function ${\rm{erf}}\left( z \right) = \frac{2}{{\sqrt \pi  }}\sum\limits_{n = 0}^\infty  {\frac{{{{\left( { - 1} \right)}^n}}}{{n!\left( {2n + 1} \right)}}} {z^{2n + 1}}$ to approximately calculate the outage probability. Hence, we derive the equation \eqref{clm2} as
\begin{align}\label{clm2}
P_{out,rfl}^{}\left( x \right)=& \sum\limits_{n = 0}^\infty  {\frac{{4{{\left( { - 1} \right)}^n}}}{{n!\sqrt \pi  \left( {2n + 1} \right){{\left( {2\eta _{eq}^{rf}} \right)}^{\frac{{2n + 1}}{2}}}}}} \sum\limits_{r = \{ 1,3, \cdots ,2n + 1\} }^{2n + 1} {\binom{2n+1}{r}} \notag\\
&\times{\left( {\bar h_{eq}^{rfl}} \right)^{2n + 1 - r}}{\left( {\frac{{{\Upsilon _{\max }}d_{BR}^{{\alpha _t}}}}{{{P_t}{C_{BR}}C_{RU}^{rfl}}}} \right)^{\frac{r}{2}}}\int_0^R {\frac{{{x^{\frac{{{\alpha _t}r}}{2} + 1}}}}{{{R^2}}}} dx,
\end{align}
and after calculating the integration $\int_0^R {\frac{{{x^{\frac{{{\alpha _t}r}}{2} + 1}}}}{{{R^2}}}} dx = \frac{{{R^{\frac{{{\alpha _t}r}}{2}}}}}{{\frac{{{\alpha _t}r}}{2} + 2}}$, we obtain the final expressions.

\vspace{-0.3cm}
\section*{Appendix~D: Proof of Theorem~\ref{cfml}} \label{Appendix:D}
\renewcommand{\theequation}{D.\arabic{equation}}
\setcounter{equation}{0}

Firstly, we substitute the CDF of the Gamma distribution, \eqref{channel32}, into the definition of the outage probability of the reflecting user, \eqref{Poutdef_rfl}. Hence, we obtain the integration as
\begin{align}
P_{out,rfl}^{}\left( x \right){\rm{ = }}\frac{{\rm{2}}}{{\Gamma \left( \alpha  \right){R^{\rm{2}}}}}\int_0^R {x\gamma \left( {\alpha ,\frac{{{\Upsilon _{\max }}d_{BR}^{{\alpha _t}}{x^{{\alpha _t}}}}}{{{P_t}{C_{BR}}C_{RU}^{rfl}{\beta _{rfl}}\beta }}} \right)} dx.
\end{align}

We then exploit the Taylor series the expand the incomplete Gamma function as $\gamma \left( {\alpha ,\beta } \right) = \sum\limits_{n = 0}^\infty  {\frac{{{{\left( { - 1} \right)}^n}{\beta ^{\alpha  + n}}}}{{n!\left( {\alpha  + n} \right)}}} $. In this way, we further calculate the equation above as the following
\begin{align}
P_{out,rfl}^{}\left( x \right) = \frac{{\rm{2}}}{{\Gamma \left( \alpha  \right){R^{\rm{2}}}}}\sum\limits_{n = 0}^\infty  {\frac{{{{\left( { - 1} \right)}^n}}}{{n!\left( {\alpha  + n} \right)}}{{\left( {\frac{{{\Upsilon _{\max }}d_{BR}^{{\alpha _t}}}}{{{P_t}{C_{BR}}C_{RU}^{rfl}{\beta _{rfl}}\beta }}} \right)}^{\alpha  + n}}} \int_0^R {{x^{{\alpha _t}\left( {\alpha  + n} \right) + 1}}} dx.
\end{align}

Finally, we derive the integration $\int_0^R {{x^{{\alpha _t}\left( {\alpha  + n} \right) + 1}}dx = } \frac{{{R^{{\alpha _t}\left( {\alpha  + n} \right) + 2}}}}{{\left[ {{\alpha _t}\left( {\alpha  + n} \right) + 2} \right]}}$ and obtain the final answers.

\begin{spacing}{0.93}
\bibliographystyle{IEEEtran}
\bibliography{mybib}
\end{spacing}
\end{document}